\documentclass[twocolumn]{aastex631}

\usepackage{amsmath}
\usepackage{bm}
\usepackage{booktabs}
\usepackage{CJKutf8}

\usepackage{soul}

\newcommand{\new}[1]{#1}
\newcommand{\newtwo}[1]{#1}

\begin{document}

\title{Filament formation due to diffusive instabilities in dusty protoplanetary disks}
\shorttitle{Diffusive Instabilities}
\shortauthors{Gerbig et al.}

\begin{CJK*}{UTF8}{bsmi}
\correspondingauthor{Konstantin Gerbig}
\email{konstantin.gerbig@yale.edu}

\author[0000-0002-4836-1310]{Konstantin Gerbig}
\affiliation{Department of Astronomy, Yale University, New Haven, CT 06511, USA}

\author[0000-0002-8597-4386]{Min-Kai Lin (林明楷)}
\affiliation{Institute of Astronomy and Astrophysics, Academia Sinica, Taipei 10617, Taiwan}
\affiliation{Physics Division, National Center for Theoretical Sciences, Taipei 10617, Taiwan}

\author[0000-0002-0496-3539]{Marius Lehmann}
\affiliation{Institute of Astronomy and Astrophysics, Academia Sinica, Taipei 10617, Taiwan}

\begin{abstract}
We report the finding of a new, local diffusion instability in a protoplanetary disk, which can operate in a dust fluid, subject to mass diffusion, shear viscosity, and dust-gas drag, provided diffusivity, viscosity, or both decrease sufficiently rapidly with increasing dust surface mass density. 
We devise a vertically averaged, axisymmetric hydrodynamic model to describe a dense, mid-plane dust layer in a protoplanetary disk.
The gas is modeled as a passive component, imposing an effective, diffusion-dependent pressure, mass diffusivity, and viscosity onto the otherwise collisionless dust fluid, via turbulence excited by the gas alone, or dust and gas in combination.
In particular, we argue that such conditions are met when the dust-gas mixture generates small-scale turbulence through the streaming instability, as supported by recent measurements of dust mass diffusion slopes in simulations.
We hypothesize that the newly discovered instability may be the origin of filamentary features, almost ubiquitously found in simulations of the streaming instability.
In addition, our model allows for growing oscillatory modes, which operate in a similar fashion as the axisymmetric viscous overstability in dense planetary rings. However, it remains speculative if the required conditions for such modes can be met in protoplanetary disks.
\end{abstract}

\keywords{}

\section{Introduction} \label{sec:intro}

Protoplanetary disks are the birthplaces of planets. One key stage within the core-accretion scenario for planet formation involves the conversion of small dust particles into km-sized planetesimals. The formation of planetesimals is associated with a multitude of challenges. Specifically, coagulational growth is thought to be inhibited at around meter sizes by both radial drift and fragmentation of small solids \citep{Birnstiel2012, Blum2018}.

For the last two decades, the attention has therefore been directed towards gravitational contraction of sufficiently massive disk regions, particle filaments, or local over-densities. Since the required dust-to-gas ratio are super-solar, one must invoke additional processes that can effectively concentrate dust particles. This includes secular gravitational instability \citep{Ward2000, Youdin2011, Takahashi2014, Tominaga2019, Tominaga2020, Tominaga2023}, particle traps such as pressure maxima \citep{Onishi2017, Shibake2020, Xu2022}, turbulent concentration \citep{Chambers2010, Hartlep2020},  and dust-gas drag instabilities \citep{Johansen2015, Schaefer2017, Gerbig2020, Gerbig2023}, the most prominent of which is the so-called streaming instability \citep{Youdin2005, Jaquet2011, Squire2018}. 

In its linear phase, the streaming instability utilizes the relative equilibrium velocity between dust and gas, in the classical picture induced by the background gas pressure gradient, to drive exponentially-growing modes \citep{Youdin2005}. The streaming instability saturates, after a few dynamic timescales, into a quasi-steady-state characterized by turbulent particle density and velocity fluctuations \citep{Johansen2007}. Eventually, this system  self-organizes into azimuthally elongated filaments which can drift inwards and merge \new{\citep[see e.g.,][]{Yang2014, Li2018, Li2021}}. 
The above three-step evolution has been readily observed in 3D shearing-box simulations of both vertically stratified and unstratified protoplanetary disks where drag and dust feedback is included.

The formation of planetesimals within the streaming instability framework requires the additional component of dust self-gravity, and albeit a priori not obvious, is thought to occur during the streaming instability's non-linear phase, either before or after the emergence of the over-dense filaments. The non-linear phase has been investigated numerically on numerous occasions. Specifically, \citet[][]{Schreiber2018} found in two-dimensional simulations that dust diffusivities tend to decrease with dust-to-gas ratio. This behavior was also seen in three-dimensional, stratified simulations by \citet[][]{Gerbig2023}, and is typically attributed to the particles carrying too much collective inertia to be effectively diffused away by residual gas turbulence, and conversely, the back-reaction of the particles' inertia onto the gas may lead to a decrease in diffusion with increasing dust-to-gas ratio. An alternative picture views particle diffusion similar to gas pressure (a model we aim to discuss thoroughly in this paper): a region of high diffusion expels particles towards regions of low diffusion.

Either way, the implication of \new{particle diffusion decreasing with increasing dust density} has hitherto not been investigated analytically in the context of the stability of dusty protoplanetary disks. \new{Previous models by \citet{Chen2020, Umurhan2020} had diffusion depend on stopping time and gas viscosity only, both of which are taken to be constant}. In this paper, we perform an instability analysis of a sheet of particles, subject to dust-gas drag forces, and mass and momentum diffusion, where the diffusion coefficients are allowed to vary with particle density. The existence of such a dependence has been established in hydrodynamic simulations by \citet{Schreiber2018, Gerbig2023}. Our treatment bears thus some similarity to hydrodynamic studies of the viscous instability \citep{LinBodenheimer1981,ward1981,Salo2010} and viscous overstability \citep{schmit1995,schmit1999,Schmidt2001,Latter2009,latter2010,lehmann2017,lehmann2019} in planetary rings. 

This paper is structured as follows. We outline our hydrodynamical model, specifically focusing on the diffusion terms and their physical relevance, as well as perform a linear perturbation analysis in Sect.~\ref{sect:modelbasics}. Next, we discuss the arising non-oscillatory and overstable modes in Sects.~\ref{sect:instability} and \ref{sect:overstability} respectively. We discuss and contextualize our results in Sect.~\ref{sect:discussion}. Lastly, Sect.~\ref{sect:summary} concludes the paper with a summary of our findings.

\section{Hydrodynamic model}
\label{sect:modelbasics}

\subsection{Diffusion and viscosity in particle-laden protoplanetary disks}

Dust diffusion has long been identified to be of immense importance for dust dynamics and consequently planetesimal formation \citep[see e.g.,][]{Cuzzi1993}. We deem it worth explicitly defining for this work the relevant properties terms and putting them into the context of previous studies related to particle diffusion. For a recent comprehensive discussion on turbulent diffusion in protoplanetary disks we refer to \citet[][]{Binkert2023}{}{}.

Generally speaking, diffusion acts to minimize free energy. In this work, we describe dust as a fluid subject to diffusion of mass, driven by a gradient in dust concentration, and momentum, driven by pressure gradients and shear stresses.

Under typical conditions, dust particles in protoplanetary disks are not collisional, and therefore do not experience collisional pressure forces.  Instead, their dynamics are influenced by their coupling to the gas, namely via their stopping time $t_\mathrm{s}$, which appears in the drag term of the momentum equation, i.e. $\bm{F}_\mathrm{d} \propto (\bm{v}- \bm{u})/t_\mathrm{s}$, where $\bm{v}$ and $\bm{u}$ are particle and gas velocity respectively. If the gas turbulence were fully characterized by the gas velocities $\bm{u}$, no additional diffusion terms would be needed in modeling dust diffusion in protoplanetary disks. Indeed, numerical simulations typically do not employ explicit diffusivity or viscosity, and instead compute diffusive effects indirectly via dust-gas interaction reflected in $\bm{u}$ \new{\citep[e.g.,][]{Yang2018, Riols2020}}. As such a treatment is often not practical for analytical progress, we employ a diffusion subgrid model and describe diffusion and viscosity due to the particles' coupling to the gas by using explicit terms in the hydrodynamical equations.

\subsection{Governing equations}
\label{sect:governing_eqs}

Specifically, in this work, we will consider an isothermal, infinitesimally thin, axisymmetric particle disk in the absence of self-gravity, embedded in a \new{gas that enters the system through diffusion, viscosity and drag}. In polar coordinates ($r,\phi$), the system is governed by the set of vertically averaged fluid equations
\begin{align}
    \label{eq:continuity_first}
    &\frac{\partial \Sigma}{\partial t} + \frac{1}{r}\frac{\partial (r\Sigma v_r)}{\partial r} = \frac{1}{r}\frac{\partial}{\partial r}\left(rD\frac{\partial \Sigma}{\partial r}\right),\\
    \begin{split}
    \label{eq:momentum_first_r}
    &\frac{\partial v_r}{\partial t} + \left(v_r - \frac{D}{\Sigma}\frac{\partial \Sigma}{\partial r}\right) \frac{\partial v_r}{\partial r} = \frac{v_\phi^2}{r} - \Omega^2 r - \frac{v_r - u_r}{t_\mathrm{s}} \\ 
    & - \frac{1}{\Sigma} \frac{\partial (c_\mathrm{d}^2\Sigma)}{\partial r}+ \frac{1}{\Sigma r} \frac{\partial}{\partial r}\left(r v_r D \frac{\partial \Sigma}{\partial r}\right) + F_r,
    \end{split}\\
    \begin{split}
           \label{eq:momentum_first_phi}
    &\frac{\partial v_\phi}{\partial t} +\left(v_r - \frac{D}{\Sigma}\frac{\partial \Sigma}{\partial r}\right) \frac{\partial v_\phi}{\partial r} = -\frac{ v_\phi}{r}\left(v_r - \frac{D}{\Sigma}\frac{\partial \Sigma}{\partial r}\right) \\ 
    &- \frac{v_\phi - u_\phi}{t_\mathrm{s}} + F_\phi. 
    \end{split}
\end{align}
Eqs. \eqref{eq:continuity_first} to \eqref{eq:momentum_first_phi} describe the dynamical evolution of the surface mass density $\Sigma$, the radial velocity $v_r$ and azimuthal velocity $v_{\phi}$, respectively, where $\Omega = \sqrt{G M_*/r^3}$ is the Keplerian angular frequency, with stellar mass $M_{*}$ and gravitational constant $G$. The continuity equation~\eqref{eq:continuity_first} takes the form of an advection-diffusion equation, with mass diffusivity for dust $D$. 
The momentum equations incorporate advection both by $v_r$ and by the diffusion flux. This leads to the modified gradient advection terms on the left hand sides of the momentum equations, as well as the modified curvature-related advection term on the right hand side of Eq.~\eqref{eq:momentum_first_phi}. In addition, the fifth term on the right hand side of Eq.~\eqref{eq:momentum_first_r} incorporates $v_r$-advection of the momentum carried in the diffusion flux itself. We refer to \citet{Tominaga2019} for a discussion on these additional advection terms associated with the diffusion flux and their implications when the full dust-gas mixture is considered, but note that they allow the total gas and dust angular momentum to be conserved.

We provide a more rigorous justification for our set of hydrodynamical equations in Appendix~\ref{sect:reynoldsavering}, using mean-field theory based on Reynolds averaging and the application of a set of plausible closure relations. Given this context, the fields $\Sigma, \bm{v}$ should be interpreted as mean fields separated in scale from the underlying small fluctuations that characterize turbulence. 

Eq.~\eqref{eq:momentum_first_r} includes the vertically averaged, effective dust pressure $P_\mathrm{d}=\Sigma c_\mathrm{d}^2$, with velocity dispersion $c_\mathrm{d}^2$ of the dust fluid. 
Since the dust is assumed to be collisionless, this effective velocity dispersion is assumed to be generated solely by the particles' coupling to the turbulence with $c_\mathrm{d}^2 \propto D$. Specifically, we follow \citet[][]{Klahr2021, Gerbig2023} and write
\begin{align}
\label{eq:particle_soundspeed}
    c_\mathrm{d}^2 = \frac{D c_\mathrm{s}^2}{t_\mathrm{s} c_\mathrm{s}^2 + D} \approx \frac{D}{t_\mathrm{s}},
\end{align}
which follows from a balance between diffusion and sedimentation. The latter approximation requires $D \ll t_\mathrm{s}c_\mathrm{s}^2$, which is the case in numerical simulations \citep[e.g.,][]{Schreiber2018, Gerbig2023}. We discuss this pressure model in Appendix~\ref{sect:dustpressuremodel}. 

Finally, we include explicit momentum diffusion terms, modeled by Navier-Stokes stress terms $F_r$ and $F_\phi$. They can be calculated via
\begin{align}
    \bm{F} = \frac{1}{\Sigma} \nabla \cdot T
\end{align}
with viscous stress tensor $T$, the components of which are
\begin{align}
\label{eq:viscous_stress_tensor}
    T_{ij} &= \nu \Sigma\left(\frac{\partial v_i}{\partial x_j}+ \frac{\partial v_j}{\partial x_i}-\frac{2}{3}\delta_{ij} \nabla \cdot \bm{v}\right),
\end{align}
where $\nu$ is the effective vertically averaged, kinematic, shear viscosity of the particle fluid. The inclusion of these shear stress terms differentiates Eq.~\eqref{eq:momentum_first_r} and Eq.~\eqref{eq:momentum_first_phi} from the dust momentum equations used by \citet[][]{Tominaga2019}.
Our model also distinguishes itself from other works concerning dusty protoplanetary disks on the scales of planetesimal formation as we consider diffusivity $D$ and viscosity $\nu$, and consequently via Eq.~\eqref{eq:particle_soundspeed} also the velocity dispersion $c_\mathrm{d}$ and dust pressure, to depend on surface mass density $\Sigma$. Specifically, we assert power law dependencies of the form 
\begin{align}
    D & \propto \left(\frac{\Sigma}{\Sigma_{0}}\right)^{\beta_\mathrm{diff}},\label{eq:D_sig}\\
    \nu & \propto \left(\frac{\Sigma}{\Sigma_{0}}\right)^{\beta_\mathrm{visc}}\label{eq:nu_sig},
\end{align}
where $\beta_\mathrm{diff}$ and $\beta_\mathrm{visc}$ are dimensionless exponents. We also introduce the corresponding dimensional slopes 
\begin{align}
    \beta_D & = \frac{\partial (D\Sigma)}{\partial \Sigma} = D(1 + \beta_\mathrm{diff}),\\
    \beta_\nu & = \frac{\partial (\nu\Sigma)}{\partial \Sigma} = \nu(1 + \beta_\mathrm{visc}).
\end{align}
At high dust-to-gas ratios, $\beta_\mathrm{D}$ has been found to be negative \citep{Schreiber2018, Gerbig2023}.  We are not aware of any numerical constraints on $\beta_\nu$ within the context of turbulent diffusion in dusty protoplanetary disks. 

\subsection{Model applicability}
\label{sect:applicability}

Equations.~\eqref{eq:continuity_first}-\eqref{eq:momentum_first_phi} 
\new{can be applied} if the combined inertia of the disk is dominated by the particle fluid, \new{for dust-to-gas volume mass density ratios $\rho_\mathrm{p}/\rho_\mathrm{g} \gtrsim 1$}. 
In this case, the presence of the gas can be reduced to perturbations that evoke drag, mass diffusion, and momentum diffusion (aka viscosity). While the analytic model itself is agnostic to the source of diffusion and viscosity, in this paper, we specifically apply it to particle layers in the midplane of a protoplanetary disk, subject to non-linear streaming instability.

Given this context, our model is restricted to radial length scales exceeding the characteristic scale of the underlying turbulence, in this case, the characteristic scale of streaming instability, which \new{for $\tau_\mathrm{s} \sim 1$,} is of order $l_\mathrm{SI} \sim \eta r$ \citep[e.g.,][]{Youdin2005, Squire2018, Gerbig2020}, where $\eta \sim 0.01$ characterizes the radial pressure gradient in the disk, and thus scales with the equilibrium relative velocity between dust and gas. \new{For smaller stopping times, this restriction is relaxed, as the characteristic scale of the linear streaming instability decreases
\citep[e.g.][Appendix D]{Lin2017}.} 
\new{Also note that in the vertical direction, this restriction is formally always satisfied as our model is vertically unstratified, implying a vanishing vertical wavenumber of all modes. Whether or not such modes are supported by an actual vertically stratified dust layer requires a stratified analysis and is thus subject to future work.}

We further \new{point out} that a fluid description for particles in protoplanetary disks as applied here, is strictly only valid if $t_\mathrm{s} < \Omega^{-1}$, since for decoupled grains the dynamical evolution of the stress tensor can, in general, not be ignored \new{\citep[see e.g.,][]{Garaud2004, Jaquet2011}}, and must be modeled using a kinetic approach. In this work, we instead assume that the `external' turbulence is able to establish a simple Newtonian stress-strain relation for the particle fluid, characterized by shear viscosity $\nu$ and isotropic velocity dispersion $c_\mathrm{d}$, as discussed in Sect.~\ref{sect:governing_eqs} and Appendix~\ref{sect:reynoldsavering}. This is to some extent similar to the effect of mutual particle collisions in planetary rings, which indeed, if frequent enough, are known to establish a Newtonian stress-strain relation of the particle flow (e.g. \citealt{stewart1984,shu1985c}). However, streaming instability turbulence, which is the main application of our model, is not expected to occur for large stopping times, which withdraws the physical justification for this assumption (at least given this context). In addition, mutual collisions, which are ignored in our model, may in principle become relevant if the velocity dispersion $c_\mathrm{d} \propto t_\mathrm{s}^{-1/2}$ becomes sufficiently small. 

Despite these limitations, we will also present and discuss results assuming larger particles with $t_\mathrm{s} > \Omega^{-1}$. In doing so, we retain a concrete connection to the viscous instability and overstability in planetary rings. Also, if large grains in protoplanetary disks do experience momentum and mass diffusion by some means (see discussion in Sect.~\ref{sect:disc_overstability}), our model may still provide useful insights, despite lacking the stress tensor evolution contained in a kinetic approach.

\subsection{Linearized equations and dispersion relation}

We adopt a local, co-rotating Cartesian reference frame at distance $R$ from the star such that $(x,y) = (r-R,R(\phi - \Omega t))$ and $v_x = v_r, v_y = v_\phi - R\Omega$. We then perturb the system around a background state such that $\Sigma = \Sigma_0 + \Sigma^\prime, v_x = v_x^\prime, v_y = -q\Omega x + v_y^\prime$ with $ \Sigma_0 = \mathrm{const.}$, and linearize in perturbed quantities. \new{Following Appendix B in \citet{Klahr2021}, we neglect the perturbed gas velocity $\bm{u}^\prime$ such that} the linearized drag terms are $\propto -\bm{v}^\prime/t_\mathrm{s}$, \new{which is justified if the mean field quantities derived in Appendix~\ref{sect:reynoldsavering} are time-averaged over one turbulent correlation time}. \new{This assumption conveniently decouples dust and gas equations and allows us to isolate the effects of dust density-dependent turbulence alone.}

For Keplerian shear where $q=3/2$, Eqs.~\eqref{eq:continuity_first}-\eqref{eq:momentum_first_phi} \new{thus} become 
\begin{align}
\label{eq:continuity_linearized}
    \frac{1}{\Sigma_0}\frac{\partial \Sigma^\prime}{\partial t} + \frac{\partial v_x^\prime}{\partial x} = & \frac{1}{\Sigma_0}\frac{\partial}{\partial x}\left(D\frac{\partial \Sigma^\prime}{\partial x}\right), \\
    \begin{split}
        \frac{\partial v^\prime_x}{\partial t} - 2 \Omega v_y^\prime = & - \frac{v^\prime_x}{t_\mathrm{s}} \\ & -\frac{1}{\Sigma_0 t_\mathrm{s}}\frac{\partial (D\Sigma)}{\partial \Sigma}\frac{\partial \Sigma^\prime}{\partial x} + \nu \frac{4}{3}\frac{\partial^2 v^\prime_x}{\partial x^2},
    \end{split}\\
    \begin{split}
    \label{eq:phi_momentum_linearized}
        \frac{\partial v^\prime_y}{\partial t} + \frac{\Omega v_x^\prime}{2} = &  -\frac{v^\prime_y}{t_\mathrm{s}}  + \frac{\Omega}{2} \frac{D}{\Sigma_0}\frac{\partial \Sigma^\prime}{\partial x} \\ & + \nu \frac{\partial^2 v^\prime_y}{\partial x^2} - \frac{3\Omega}{2\Sigma_0} \frac{\partial (\nu\Sigma )}{\partial \Sigma}\frac{\partial \Sigma^\prime}{\partial x},
    \end{split}
\end{align}
This set of linearized equations is novel in that it includes a Navier-Stokes viscosity for the particle fluid, relates the particle pressure to diffusion and stopping time via Eq.\eqref{eq:particle_soundspeed}, which produces the second term on the r.h.s. of the radial momentum equation, and takes into account the dependence of diffusion and viscosity on the particle surface mass density, as motivated by simulations of \citet{Schreiber2018} and \citet{Gerbig2023}. As a consequence, the radial and azimuthal momentum equations respectively contain the slope of diffusion and viscosity with respect to particle surface mass density. Depending on the slope, these terms can act both stabilizing or destabilizing on perturbations to the equilibrium state defined above, as we discuss below. The mass diffusion term in the continuity equation, the  terms $ \propto \partial/\partial x^2 $ (assuming $\nu>0$), and the term describing advection of background shear by the diffusion flux (second term on r.h.s of Eq.~\eqref{eq:momentum_first_phi}) are always stabilizing.  Note, that this diffusion flux term is the only term from the four angular momentum conserving terms in Eqs.~\eqref{eq:momentum_first_r} and \eqref{eq:momentum_first_phi} added by \citet{Tominaga2019} that survives linearization. 
Lastly, the drag terms have a stabilizing effect. In our analysis, we include both radial and azimuthal drag terms, which is in contrast to \citet{Klahr2021} who drop the azimuthal drag term. 

We proceed by introducing axisymmetric modes of the form 
\begin{align}
\label{eq:perturbuation_def}
    f^\prime = \Re\left(\hat{f} e^{-ikx + n t}\right),
\end{align}
with complex frequency $n$ and (radial) wavenumber $k$. We take $k>0$ without loss of generality. Modes grow and decay for $\Re{(n)} > 0$ and $\Re{(n)} < 0$ respectively, and $\Im(n)$ corresponds to the oscillation frequency, the sign of which sets the wave travel direction. We get
\begin{align}
    n \frac{\hat{\Sigma}}{\Sigma_0} =  & ik\hat{v}_x - D k^2 \frac{\hat{\Sigma}}{\Sigma_0},\\
   n \hat{v}_x - 2\Omega \hat{v}_y =  & -\frac{\hat{v}_x}{t_\mathrm{s}} + \frac{ik}{t_\mathrm{s}} \beta_D \frac{\hat{\Sigma}}{\Sigma_0} - \frac{4}{3} \nu k^2 \hat{v}_x,\\
    \begin{split}
        n \hat{v}_y + \frac{\Omega}{2}\hat{v}_x  = &  - \frac{\hat{v}_y}{t_\mathrm{s}} - \frac{ik}{2}D\Omega \frac{\hat{\Sigma}}{\Sigma_0} \\
        &-   \nu k^2 \hat{v}_y  + \frac{3}{2} i k \Omega \beta_\nu \frac{\hat{\Sigma}}{\Sigma_0}.
    \end{split}
\end{align}

This system is solved by a cubic dispersion relation of the form 
\begin{align}
\label{eq:disprel_n}
    n^3 + n^2 a_2  +n a_1 + a_0 = 0,
\end{align}
with coefficients
\begin{align}
    a_2 =& \left(\frac{7}{3}\nu  + D\right)k^2 + \frac{2}{t_\mathrm{s}}, \\
    \begin{split}
        a_1 =& \left(\frac{7}{3}D\nu + \frac{4}{3}\nu^2\right)k^4 \\
         &+\left(\frac{2D}{t_\mathrm{s}} + \frac{7}{3}\frac{\nu}{t_\mathrm{s}} + \frac{\beta_D}{t_\mathrm{s}} \right)k^2 \\
         &+\frac{1}{t_\mathrm{s}^2} + \Omega^2
    \end{split}\\
    \begin{split}
        a_0 = & \left(\frac{4}{3}D\nu^2\right)k^6 + \left(\frac{7}{3}\frac{D\nu}{t_\mathrm{s}} + \frac{\nu}{t_\mathrm{s}}\beta_D\right)k^4 
       \\ & +\left(\frac{D}{t_\mathrm{s}^2} + \frac{\beta_D}{t_\mathrm{s}^2} + 3\Omega^2 \beta_\nu\right)k^2,
    \end{split}
\end{align}

\subsection{Dimensionless quantities}

It is convenient to write the dispersion relation in terms of commonly used dimensionless quantities. The orbital frequency introduces a time unit, such that we write the dimensionless stopping time as
\begin{align}
    \tau_\mathrm{s} \equiv t_\mathrm{s}\Omega,
\end{align}
which, in general, is distinct from the so-called Stokes number defined as the ratio of stopping time and turbulent correlation time (also known as eddy time or integral time) \citep{Cuzzi1993, Youdin2007}. Particles with $\tau_\mathrm{s}\ll1$ are well-coupled to the gas; while $\tau_\mathrm{s}\gg1$ applies to loosely-coupled dust.

Our reference length unit is the gas pressure scale height, which is the ratio of the (gas) sound speed $c_\mathrm{s}$ and orbital frequency, $H = c_\mathrm{s}/\Omega$. We write the dimensionless wave number as $K \equiv kH$. Dimensionless versions for ground state diffusivity in Eq.\eqref{eq:D_sig} and viscosity in Eq.~\eqref{eq:nu_sig} are introduced as
\begin{align}
    \delta \equiv \frac{D}{c_\mathrm{s}H},\label{eq:delta}\\
    \alpha \equiv \frac{\nu}{c_\mathrm{s}H}\label{eq:alpha},
\end{align}
where we use the same nomenclature as the well-known Shakura Sunyaev $\alpha$-parameter \citep{Shakura1973}, albeit in our work, $\alpha$ describes the effective viscosity of the dust fluid, and not the viscosity of the gas that is often adopted to model angular momentum transport in protoplanetary disks. The corresponding power law slopes of diffusivity and viscosity were defined in Eqs.~\eqref{eq:D_sig} and \eqref{eq:nu_sig}, i.e. $\beta_\mathrm{diff} = \partial \ln \delta/ \partial \ln \Sigma$ and $\beta_\mathrm{visc} = \partial \ln \alpha/\partial\ln  \Sigma$ respectively. Typical values for diffusivity and its slope are $ 10^{-5} \lesssim \delta \lesssim 10^{-4}$, and $-3 \lesssim \beta_\mathrm{diff} \lesssim -1 $ at high dust-to-gas ratios \citep{Schreiber2018, Gerbig2023}. Appropriate constraints on particle viscosity $\alpha$ are less clear. In resistive simulations of the magneto-rotational instability \citep{Balbus1991},  \citet{Yang2018} measured a shear viscosity of $\alpha \sim 10^{-4}$ in the gas. However, that value includes (albeit presumably small)  contributions from Maxwell stresses, and constitutes an average value throughout the particle layer. We are not aware of any other applicable constraints on $\alpha$ in the particle layer, or any measurements of $\beta_\mathrm{visc}$. 

We define the hydrodynamic Schmidt number as the ratio of viscosity and mass diffusion coefficient for our particle fluid, i.e.
\begin{align}
    \mathrm{Sc} \equiv \frac{\alpha}{\delta},
\end{align}
which is analogous to the definition used for pure gas in protoplanetary disks by \citet[][]{Johansen2005, Carballido2006}{}{}. As pointed out in \citet[][]{Youdin2007}{}{}, in the context of particles in protoplanetary disks, the Schmidt number suffers from being over-subscribed, as it is also used to describe the ratio of gas viscosity and particle diffusivity \citep{Cuzzi1993, Schreiber2018, Binkert2023}. Here, we are primarily interested in the dust component, hence we assign $\mathrm{Sc}$ to characterize the relative importance of particle viscosity and particle diffusivity, both of which stem from the particles' coupling to gas turbulence.

Lastly, we notice that for $\delta > 0$, the dispersion relation includes a degeneracy in wave number and diffusivity, such that we can define 
\begin{align}
    \xi \equiv \delta K^2 = \frac{D k^2}{\Omega},
\end{align}
Note, that since the herein utilized description for dust-pressure is only appropriate for $\tau_\mathrm{s}\gg \delta$, we equivalently require $\xi \ll \tau_\mathrm{s} K^2$.  As noted in Sect.~\ref{sect:applicability}, our model is only appropriate for scales larger than the scale of the under-lying turbulence. We thus require, $\xi \lesssim  4\pi^2 \delta / ((H/r)^2 \eta^2$). For typical values of $H/r = 0.1$, $\eta \sim 1\%$, and for fiducial diffusivities of $\delta \sim 10^{-5} - 10^{-4}$, this corresponds to maximum $\xi$ of order unity. This is why, for the remainder of this study, we will mostly be concerned with the long-wavelength limit where $\xi < 1$.

The dimensionless complex frequency is denoted as 
\begin{align}
    \sigma \equiv \frac{n}{\Omega} \equiv \gamma + i\omega,
\end{align}
where $\gamma = \Re{(\sigma)}$ is the growth rate and $\omega = \Im{(\sigma)}$ is the oscillation frequency. The dimensionless version of the cubic dispersion relation in Eq.~\eqref{eq:disprel_n} is given by
\begin{align}
\label{eq:disprel_sigma}
    \sigma^3 +\sigma^2 A_2 + \sigma A_1 + A_0 = 0,
\end{align}
with
\begin{align}
    A_2 =& \left(\frac{7}{3}\mathrm{Sc}  + 1\right)\xi + \frac{2}{\tau_\mathrm{s}}, \\
    \begin{split}
        A_1 =& \frac{\mathrm{Sc}}{3}\left(7 + 4\mathrm{Sc}\right)\xi^2 \\
         &+\left(\frac{7}{3}\mathrm{Sc}+3+ \beta_\mathrm{diff}\right)\frac{\xi}{\tau_\mathrm{s}}+ \frac{1}{\tau_\mathrm{s}^2} + 1,
    \end{split}\\
    \begin{split}
    \label{eq:A0}
    A_0 = &\frac{4}{3}\mathrm{Sc}^2\xi^3 + \frac{\mathrm{Sc}}{\tau_\mathrm{s}}\left(\frac{10}{3} + \beta_\mathrm{diff}\right)\xi^2 \\ &+\left[\frac{1}{\tau_\mathrm{s}^2} \left(2+\beta_\mathrm{diff}\right) + 3\mathrm{Sc} \left(1 +\beta_\mathrm{visc}\right)\right]\xi,
    \end{split}
\end{align}
where $A_0, A_1, A_2$ are real coefficients. This implies that $A_0 < 0$ is a sufficient condition for non-oscillatory instability, as then there is at least one real positive root of Eq.~\eqref{eq:disprel_sigma}.
\new{The full, complex dispersion relation can also be cast into two equations for the growth rate and oscillation frequency, which need to hold independently:
\begin{align}
\label{eq:real_part_overstab}
    \omega^3 - 3\omega \gamma^2 - 2\omega \gamma A_2 - \omega A_1 &= 0, \\
    \gamma^3 - 3\omega^2 \gamma - \omega^2A_2 + \gamma^2 A_2 +\gamma A_1 + A_0 &= 0.
\label{eq:imag_part_overstab}
\end{align}
}

\section{Diffusive Instability}
\label{sect:instability}

This section investigates the diffusive instability associated with the real roots of the dispersion relation in Eq.~\eqref{eq:disprel_sigma}. Consider, however, first the case with zero diffusivity and viscosity, i.e. $\mathrm{Sc} = 0$, $\xi =0$. In this limit, the above dispersion relation reduces to 
\begin{align}
    \sigma\left(\sigma^2 + \frac{2}{\tau_\mathrm{s}}\sigma + \frac{1}{\tau_\mathrm{s}^2} + 1\right) = 0,
\end{align}
which is solved by damped, epicyclic waves with oscillation frequency $\Omega$ and negative growth rate $-t_\mathrm{s}^{-1}$. The purely real root is the static null solution with $\sigma = \gamma = \omega = 0$. It is this mode that will get destabilized by diffusion and/or viscosity, as we discuss in the following. As this section is only concerned with the non-oscillatory, purely real solution, we set $\omega = 0$, and replace $\sigma = \gamma$. We will redirect our attention to oscillatory modes in Sect.~\ref{sect:overstability}.

\subsection{Inviscid case}

\begin{figure}
    \centering
    \includegraphics[width = \linewidth]{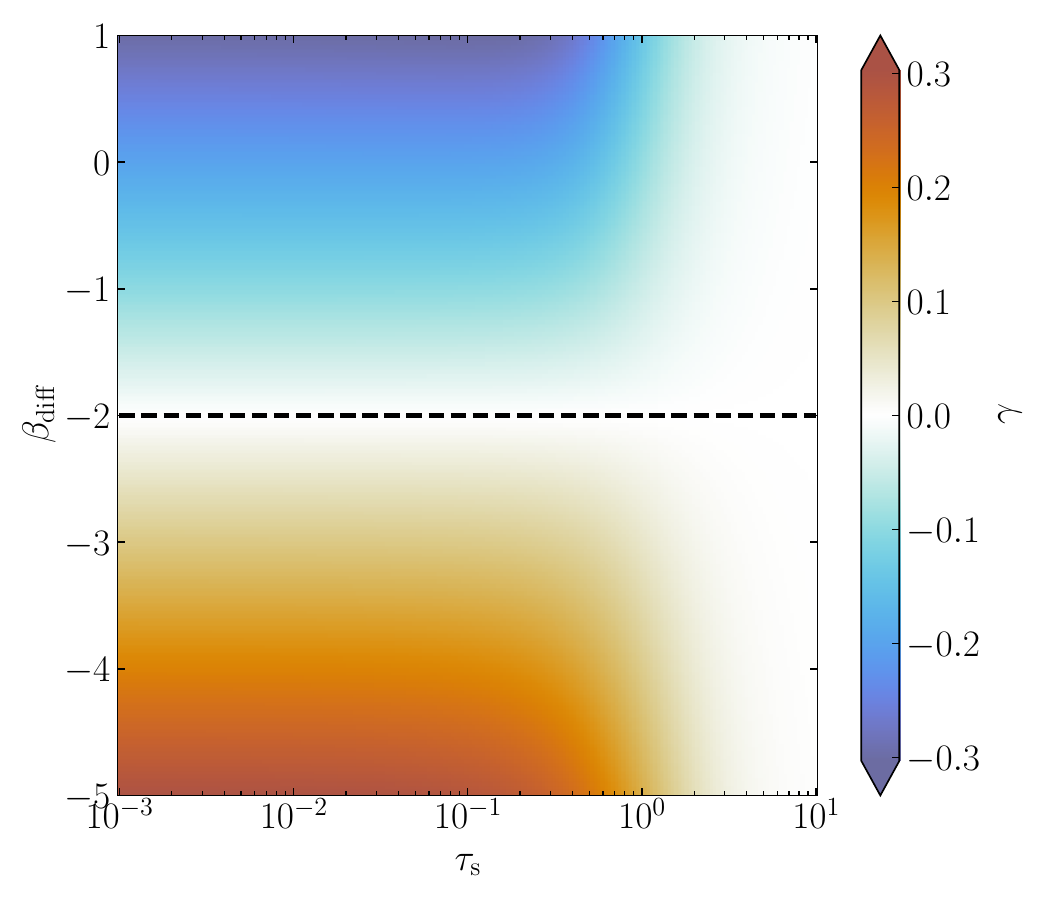}
    \caption{Growth rates for the inviscid ($\mathrm{Sc} = 0$) diffusive instability driven by the diffusion-dependent pressure for $\xi =0.1$. The black dashed line corresponds to Eq.~\eqref{eq:inviscid_diffusive_instability_criterion}, below which the instability can operate, and above which, perturbations are exponentially damped.}
    \label{fig:inviscid_diff_inst_xi0.1}
\end{figure}

\begin{figure}
    \centering
    \includegraphics[width = \linewidth]{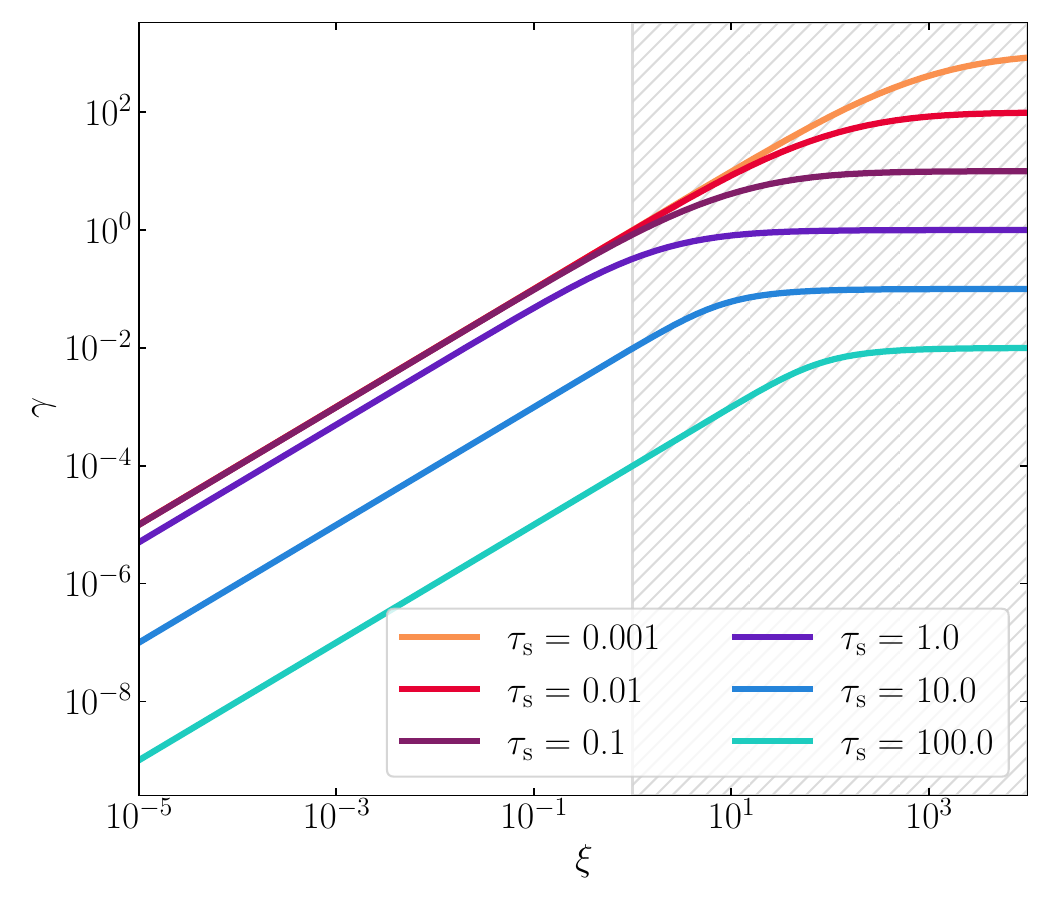}
    \caption{Growth rates for the inviscid ($\mathrm{Sc} = 0$) diffusive instability driven by the diffusion-dependent pressure for $\beta_\mathrm{diff} = -3$ and various stopping times. The hatched region corresponds to $\xi > 1$, for which our model seizes to be appropriate. 
    }
    \label{fig:inviscid_diff_inst_growthrates}
\end{figure}

We first investigate the purely diffusive case, where $\mathrm{Sc} = 0$. In this limit the dispersion relation can be written as
\begin{align}
\begin{split}
\label{eq:disprel_invsicid}
    \gamma^3 + &\left(\xi + \frac{2}{\tau_\mathrm{s}}\right)\gamma^2 
    + \left[\left(3 + \beta_\mathrm{diff}\right)\frac{\xi}{\tau_\mathrm{s}} + \frac{1}{\tau_\mathrm{s}^2} + 1\right]\gamma \\ &
    + \frac{\xi}{\tau_\mathrm{s}^2}\left(2 + \beta_\mathrm{diff}\right) = 0,
\end{split}
\end{align}
which leads to unconditional instability if
\begin{align}
\label{eq:inviscid_diffusive_instability_criterion}
    \beta_\mathrm{diff} < - 2.
\end{align}
This value is of order the power law slopes in \citet{Schreiber2018, Gerbig2023}.

For small growth rates $\gamma \ll 1$, we find from Eq.~\eqref{eq:disprel_invsicid},
\begin{align}
    \gamma = - \frac{ \xi(2 + \beta_\mathrm{diff})}{1 + \xi \tau_\mathrm{s}(3 + \beta_\mathrm{diff}) + \tau_\mathrm{s}^2},
\end{align}
which for small stopping times  equals
\begin{align}
\label{eq:inviscid_diffusive_inst_limit_sig}
    \gamma \simeq  - \xi \left(2 + \beta_\mathrm{diff}\right) \quad ,(\tau_\mathrm{s}\ll 1).
\end{align}
On the other hand. for large stopping times with $\tau_\mathrm{s} \gg 1$, growth rates are uninterestingly small. Fig.~\ref{fig:inviscid_diff_inst_xi0.1} shows the growth rate for the inviscid diffusive instability for choices of slope $\beta_\mathrm{diff}$ and stopping time $\tau_\mathrm{s}$ for modes with $\xi = 0.1$, which for a fiducial $\delta = 10^{-5}$ corresponds to $K = 100$. 

Fig.~\ref{fig:inviscid_diff_inst_growthrates} shows the growth rates of the inviscid diffusive instability for $\beta_\mathrm{diff} = -3$ for various stopping times. 
The system is unstable for all $\xi$ although growth rates become uninterestingly small for large stopping times. 
In fact, we note that inviscid diffusive instability is not damped at small scales, but displays fastest growth rates at scales $\xi > 1$, which is where our model breaks down, as indicated by the hatched region in Fig.~\ref{fig:inviscid_diff_inst_growthrates}. 

The physical origin of the inviscid diffusive instability lies in the dust pressure term in Eq.~\eqref{eq:momentum_first_r}. For $\beta_\mathrm{diff} < 0$, this pressure term acts destabilizing and accelerates particles towards annuli of high density and low diffusivity. If $\beta_\mathrm{diff} < -2$, this pressure forcing can overcome the stabilizing mass diffusion term in Eq.~\eqref{eq:continuity_first} and the drag terms in the momentum equations. This increases the density, in the process further decreasing diffusion, resulting in positive feedback and instability. While in the large stopping time limit, the drag terms vanish, the destabilizing pressure term does too. The only remaining terms are the mass diffusion term in the continuity equation, and the diffusion flux term in Eq.~\eqref{eq:phi_momentum_linearized}, both of which act to repel particles away from density maxima. Hence, only for small stopping times are the growth rates of this instability appreciably fast.

\subsection{Sc = 1 case}

\begin{figure}
    \centering
    \includegraphics[width = \linewidth]{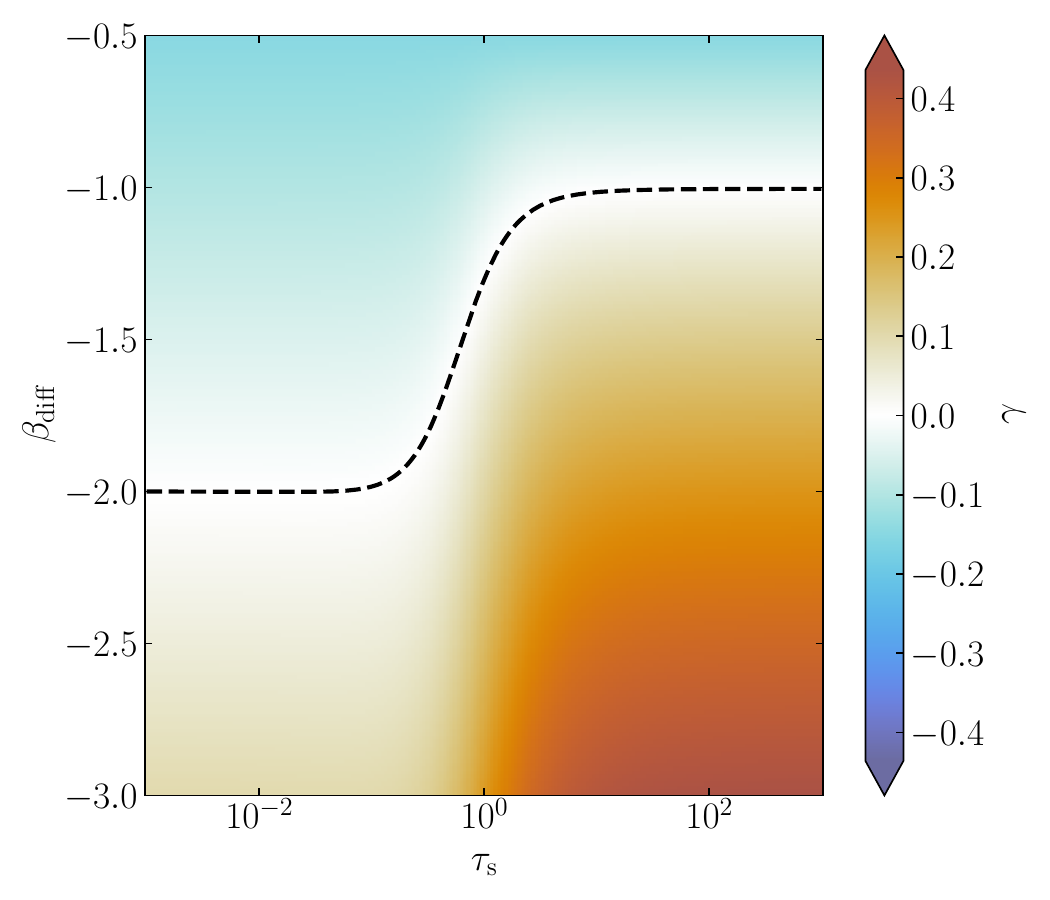}
    \caption{Like Fig.~\ref{fig:inviscid_diff_inst_xi0.1}. Growth rates for the viscous-diffusive instability for $\xi =0.1$, assuming equal mass and momentum diffusion with $\mathrm{Sc} = 1$ and $\beta_\mathrm{diff} = \beta_\mathrm{visc}$. The black curve corresponds to Eq.~\eqref{eq:Sc1_pureinstcrit}, below which the instability can operate, and above which  perturbations are exponentially damped. Note that while parts of the depicted parameter space allow for viscous overstability, this figure is only showing the purely real solution. See Sect.~\ref{sect:overstability}. }
    \label{fig:Sc=1_diff_inst_xi0.1}
\end{figure}

\begin{figure}
    \centering
    \includegraphics[width = \linewidth]{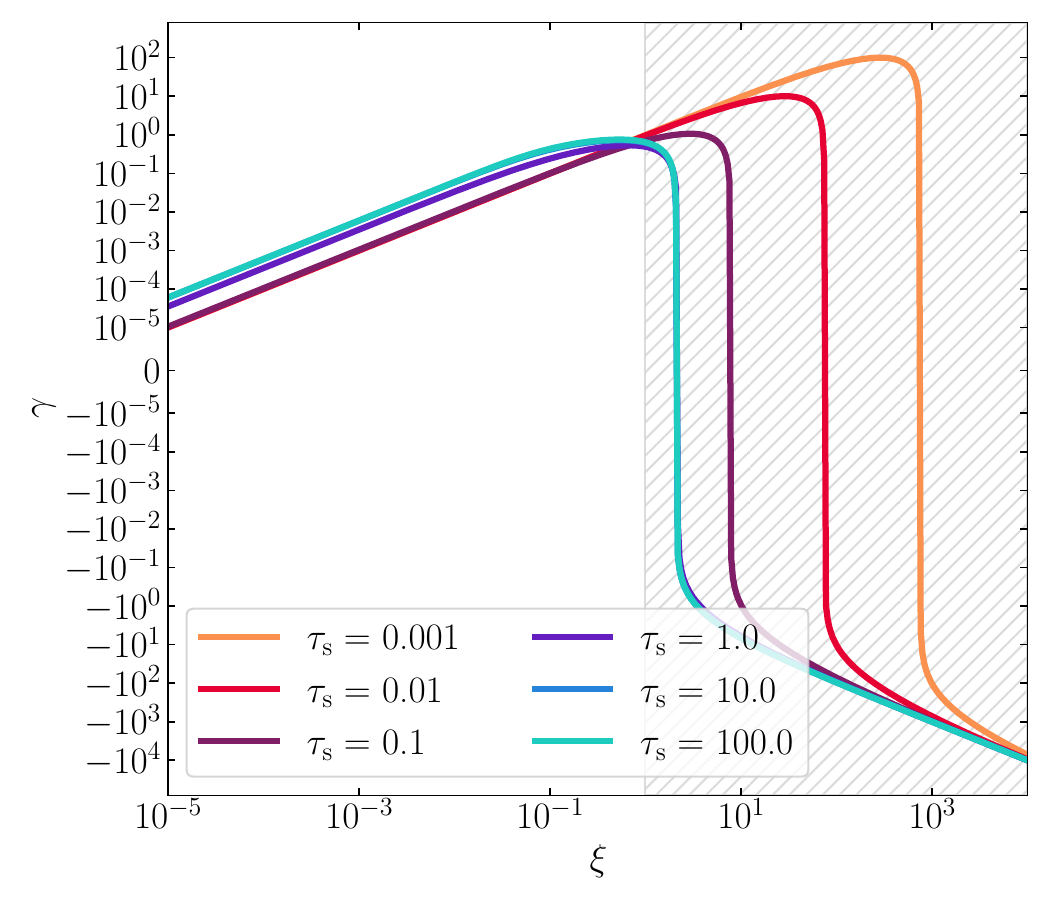}
    \caption{Like Fig.~\ref{fig:inviscid_diff_inst_growthrates}, but for the viscous-diffusive instability assuming equal mass and momentum diffusion with $\mathrm{Sc} = 1$ and $\beta_\mathrm{diff} = \beta_\mathrm{visc} = -3$, for various stopping times. The viscosity terms damp the instability on small scales, albeit only for $\xi > 1$, where we expect the model to break down. \new{Lines for $\tau_\mathrm{s} = 10$ and $\tau_\mathrm{s} = 100$ overlap.}}
    \label{fig:visc_diff_inst_growthrates}
\end{figure}

Next, we consider the case where $D = \nu$, equivalently $\mathrm{Sc} = 1$, such that the equilibrium value of momentum diffusion equals that of the mass diffusion, and also the power law slopes of diffusion and viscosity are identical, i.e. $\beta_\mathrm{visc} = \beta_\mathrm{diff}$. Under these assumptions, the dispersion relation is
\begin{align}
\begin{split}
\label{eq:Sc1_fullcubic}
       \gamma^3 &+ \left(\frac{10}{3}\xi + \frac{2}{\tau_\mathrm{s}}\right)\gamma^2 \\ &+ \left[\frac{11}{3}\xi^2 + \left(\frac{16}{3} + \beta_\mathrm{diff}\right)\frac{\xi}{\tau_\mathrm{s}}+\frac{1}{\tau_\mathrm{s}^2}+1\right]\gamma \\ &+\frac{4}{3}\xi^3 + \left(\frac{10}{3} + \beta_\mathrm{diff}\right)\frac{\xi^2}{\tau_\mathrm{s}} \\ &+ \left[3+\frac{2}{\tau_\mathrm{s}^2} + \left(3 + \frac{1}{\tau_\mathrm{s}^2}\right) \beta_\mathrm{diff}\right]\xi = 0,
\end{split}
\end{align}
Pure instability is achieved if 
\begin{align}
\label{eq:Sc1_pureinstcrit}
    \beta_\mathrm{diff} < -\frac{4\xi^2\tau_\mathrm{s}^2 + 10\xi\tau_\mathrm{s} + 9 \tau_\mathrm{s}^2 + 6}{3\xi\tau_\mathrm{s} + 9 \tau_\mathrm{s}^2 + 3}.
\end{align}
In the long wavelength limit of $\xi < 1$, we recover our finding for the inviscid case where a power law gradient $\beta_{\text{diff}}<-2$ suffices for instability, if stopping times are sufficiently small: $\tau_\mathrm{s} \ll 1$. This is because for small stopping times and at large radial length scales, the destabilizing pressure term dominates over the viscosity terms. 


For large stopping times of $\tau_\mathrm{s} \gg 1$, the explicit drag terms and the pressure term become neglible. In this case, the long wavelength limit $\xi < 1$ yields
\begin{align}
\label{eq:Sc1_visc_inst_crit}
    \beta_\mathrm{diff} \lesssim - \frac{(4\xi^2 +9)\tau_\mathrm{s} + 10\xi}{9\tau_\mathrm{s} + 3\xi} \approx - 1,
\end{align}
The diffusive instability behaves now analogously to the classical viscous instability \citep{ward1981,LinBodenheimer1981}, in that the instability is driven by the density slope of shear viscosity that appears in the azimuthal momentum equation~\eqref{eq:phi_momentum_linearized}. If the slope is sufficiently steep, there is a net flux towards density maxima that amplifies the linear perturbation. The criterion for the classical viscous instability is given by $\beta_\mathrm{visc} < -1$. 

It is easy to see in Eq.~\eqref{eq:A0}, that if $\tau_\mathrm{s}\ll 1$, the slope of mass diffusion will dominate, whereas for $\tau_\mathrm{s}\gg 1$, the slope in momentum diffusion will dominate, thus leading to diffusive instability driven by the pressure term and viscosity terms respectively. For marginally coupled particles $\tau_\mathrm{s} \sim 1$, the instability can utilize both slopes. Because of this, on large scales $\xi \lesssim 1$, if $\beta_\mathrm{diff}$ is sufficiently negative, the system is unstable regardless of stopping time.

For small growth rates $\gamma \ll 1$, and 
for small $\xi$, the real root of Eq.~\eqref{eq:Sc1_fullcubic} behaves as
\begin{align}
    \gamma &\simeq  - \frac{\left[3+2\tau_\mathrm{s}^{-2}+\left(3+ \tau_\mathrm{s}^{-2}\right)\beta_\mathrm{diff}\right]\xi}{\tau_\mathrm{s}^{-1}(16/3 + \beta_\mathrm{diff})\xi+\tau_\mathrm{s}^{-2} + 1},
\end{align}
which in the limits of well-coupled and decoupled particles equals
\begin{align}
 \gamma \simeq   \begin{cases}
			- \xi \left(2 + \beta_\mathrm{diff}\right) & \text{$\tau_\mathrm{s}\ll 1$}\\
            -3\xi(1 + \beta_{\mathrm{diff}}) & \text{$\tau_\mathrm{s}\gg 1$}
		 \end{cases}
\end{align}
respectively. We thus recover Eq.~\eqref{eq:inviscid_diffusive_inst_limit_sig}, in the limit of small stopping times.

Fig.~\ref{fig:Sc=1_diff_inst_xi0.1} shows the real root of the full cubic in Eq.~\eqref{eq:Sc1_fullcubic} for $\xi =0.1$. The black curve corresponds to Eq.~\eqref{eq:Sc1_pureinstcrit}. For a given $\beta_\mathrm{diff}$, growth rates are greater for large stopping times. Fig.~\ref{fig:visc_diff_inst_growthrates} depicts the growth rate of the viscous-diffusive instability for $\beta_\mathrm{diff} = \beta_\mathrm{visc} = -3$ for various stopping times. Unlike in the invisicd case, where if Eq.~\eqref{eq:inviscid_diffusive_instability_criterion} is satisfied, all modes $\xi$ are unstable (see Fig.~\ref{fig:inviscid_diff_inst_growthrates}), the viscous diffusive instability is damped at small scales ($\xi \gtrsim 1$) by viscosity. This regularizes the system, by prohibiting growth on arbitrarily small scales. While our model itself is not applicable to large $\xi$, one does not expect growth rates to increase without bound at ever-decreasing scales, which is an advantage of including viscosity in the model.

\subsection{General case}

Given that the numerical constraints on the effective viscosity in high-density particle mid-plane of protoplanetary disks are sparse, we finally investigate the most general case our model allows, where we remain agnostic to the value of $\mathrm{Sc}$ and retain two independent power law slopes in $\delta$ and $\alpha$. As before, pure instability is achieved for $A_0 < 0$, which for small $\xi$ results in the condition
\begin{align}
    \frac{1}{\tau_\mathrm{s}^2}\left(2+ \beta_\mathrm{diff}\right) + 3\mathrm{Sc}\left(1 + \beta_\mathrm{visc}\right) < 0,
\end{align}
In the general case, the growth rate of the diffusive instability, as well as the existence thereof in the first place, thus depends on five parameters: the stopping time $\tau_\mathrm{s}$, Schmidt number $\mathrm{Sc}$, diffusion slope $\beta_\mathrm{diff}$, viscosity slope $\beta_\mathrm{visc}$ and dimensionless wave number $\xi$.  The Schmidt number $\mathrm{Sc}$ acts as an amplification to the viscosity slope in the context of the diffusive instability.

Like in the previous two cases, well-coupled particles with $\tau_\mathrm{s}\ll 1$ result in instability if $\beta_\mathrm{diff} < - 2$, on account of the diffusive instability. On the other hand, loosely-coupled particles with $\tau_\mathrm{s}\gg 1$ are unstable if 
\begin{align}
    \beta_\mathrm{visc} < -1.
\end{align}
We recover the classical criterion for viscous instability in planetary rings \citep{ward1981,LinBodenheimer1981} .

For pure instability with small growth rates, $\gamma \ll 1$, Eq.~\ref{eq:disprel_sigma}  yields $\gamma = - A_0/A_1$, which for $\xi < 1$ results in
\begin{align}
    \gamma = - \frac{\xi\left[ \tau_\mathrm{s}^{-2}\left(2 + \beta_\mathrm{diff}\right) + 3\mathrm{Sc}\left(1+\beta_\mathrm{visc}\right)\right]}{\tau_\mathrm{s}^{-1}(7\mathrm{Sc}/3 + 3+ \beta_\mathrm{diff})\xi+\tau_\mathrm{s}^{-2} + 1}.
\end{align}
In the well-coupled limit $\tau_\mathrm{s} \ll 1$, this equals the inviscid case in Eq.~\eqref{eq:inviscid_diffusive_inst_limit_sig}. In the loosely-coupled limit with $\tau_\mathrm{s}\gg 1$, we get
\begin{align}
    \gamma = 3\mathrm{Sc}\xi(1+\beta_\mathrm{visc}) \quad ,(\tau_\mathrm{s} \gg 1).
\end{align}

\section{Overstability}
\label{sect:overstability}

\begin{figure*}
    \centering
    \includegraphics[width = \linewidth]{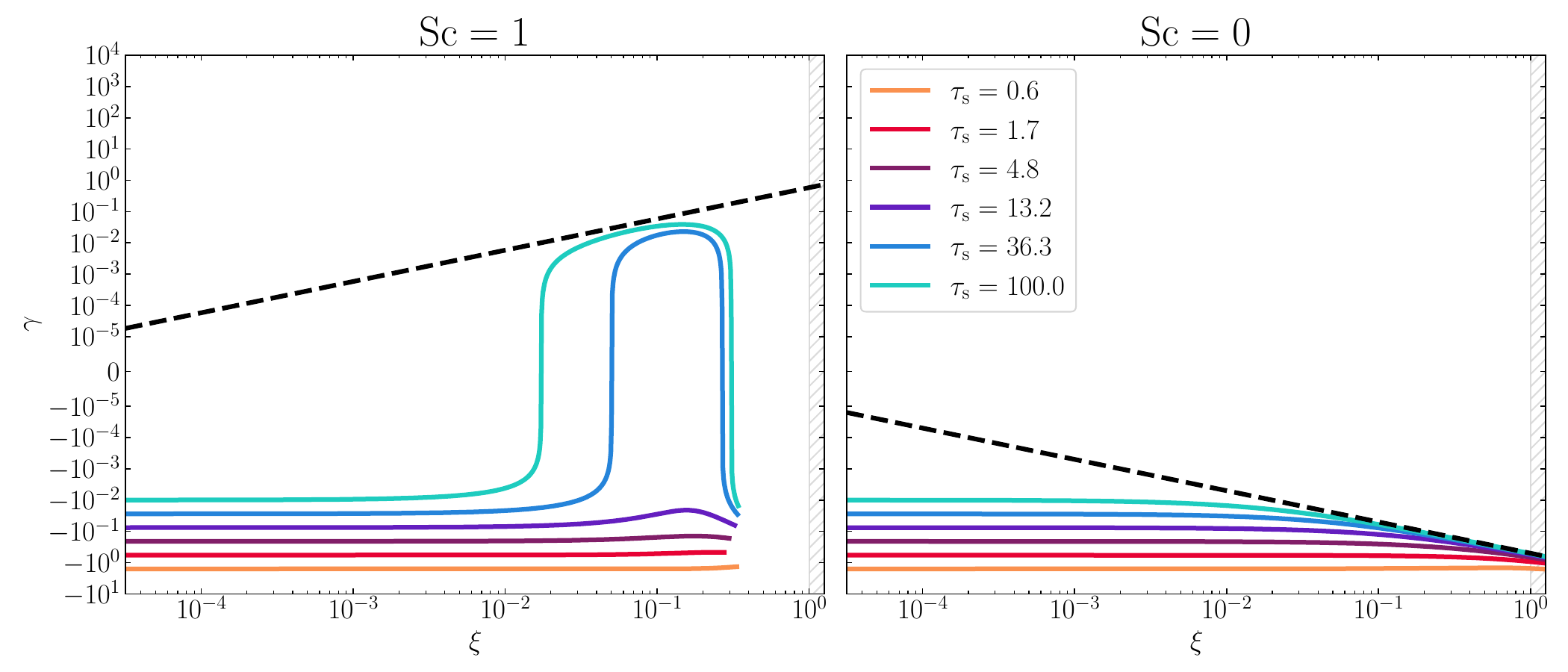}
    \caption{Growth rates for \new{diffusive overstability driven by viscosity slope ($\mathrm{Sc} = 1$, left panel) compared to the damping rates in the absence of a viscosity ($\mathrm{Sc} = 0$, right panel)} for various stopping times \new{obtained from the full dispersion relation in Eq.~\eqref{eq:disprel_sigma}}. We set $\beta_\mathrm{visc} = \beta_{\mathrm{diff}} = 1/2$. 
    The dashed lines correspond to Eq.~\eqref{eq:gamma_overstab_approx}.
    \new{The lines end once modes become non-oscillatory (and thus damped), which occurs as $\xi$ approaches the hatched region where our model is not applicable.}}
\label{fig:overstability_growth_rates}
\end{figure*}

\begin{figure}
    \centering
    \includegraphics[width = \linewidth]{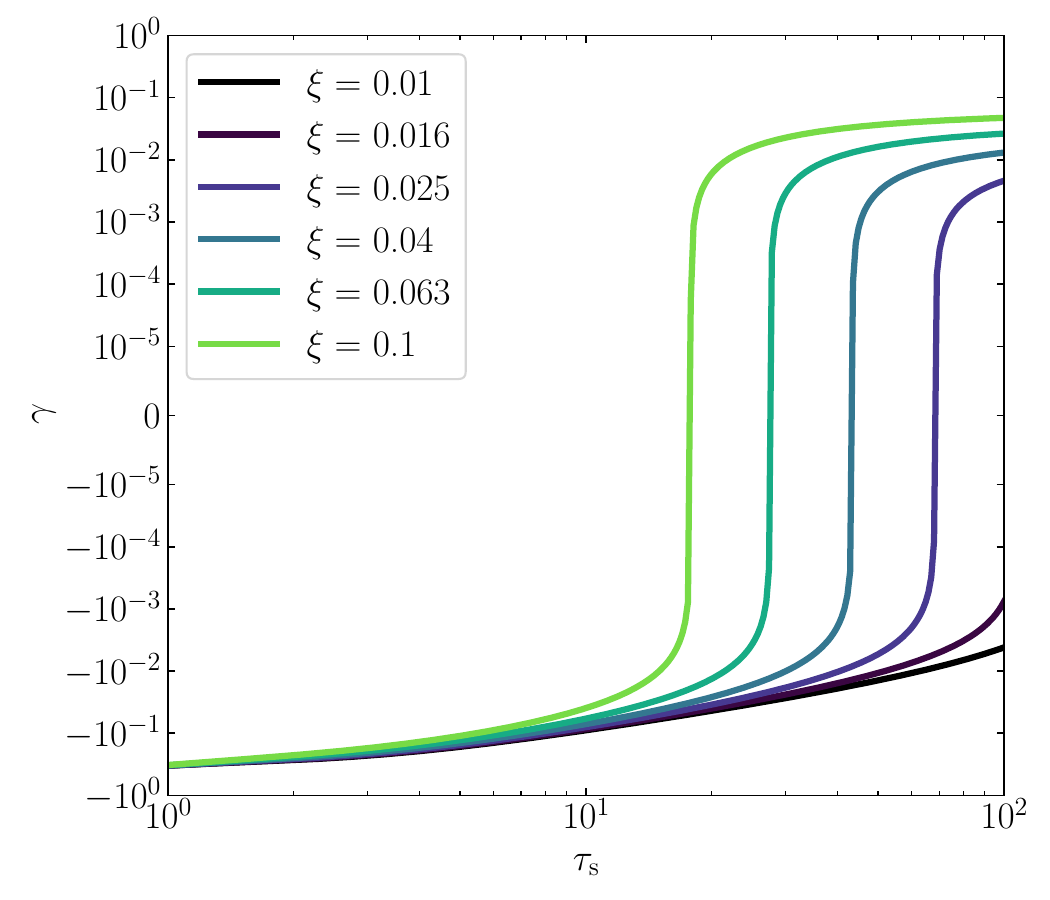}
    \caption{\new{Growth rates for diffusive overstability driven by viscosity slope ($\mathrm{Sc} = 1$) vs stopping time for various values of $\xi$; setting $\beta_\mathrm{visc} = \beta_{\mathrm{diff}} = 1/2$ following Eq.~\eqref{eq:overstab_gamma_smallxi}, which assumes $\xi \ll 1$. For the smallest scales our model applies to, and, given the chosen set of parameters, stopping times of $\tau_\mathrm{s} \gtrsim 8$ would allow for overstability.}}
    \label{fig:overstability_s growth_rates_vs_tau}
\end{figure}

We now direct our attention to \new{overstable} modes with nonzero $\omega$. \new{Figure.~\ref{fig:overstability_growth_rates} shows growth rates of such oscillatory modes as obtained from the full dispersion relation, and demonstrates that large stopping times are required to achieve growing modes.} \newtwo{We caution that in this regime, the fluid approximation for the dust grains that underlies our model starts to break down (also see Sects.~\ref{sect:applicability}, \ref{sect:disc_overstability} as well as Appendix \ref{sect:reynoldsavering}). We proceed with the analysis for completeness.} 

For oscillating solutions with $\omega \neq 0$, Eq.~\eqref{eq:real_part_overstab} implies 
\begin{align}
\label{eq:oscillation_freq}
    \omega^2 = 3\gamma^2 + 2\gamma A_2 + A_1,
\end{align}
which is positive definite \new{for $\xi \ll 1$ as seen in Fig.~\ref{fig:overstability_growth_rates}}. 
When inserted into (\ref{eq:imag_part_overstab}), this results  in a cubic for the growth rate of the wave, i.e.
\begin{align}
\label{eq:overstab_cubic}
    8\gamma^3 + 8A_2\gamma^2 + 2\left(A_1 + A_2^2\right)\gamma + A_1 A_2 - A_0 = 0.
\end{align}
For small growth rates $\gamma \ll 1, A_1, A_2$, the root  is
\begin{align}
    \gamma \simeq - \frac{A_1 A_2 - A_0}{2(A_1 + A_2^2)}.
    \label{eq:Gamma_small_growth_rates}
\end{align}
For $\xi\ll1$, we have
\begin{align}
\label{eq:overstab_gamma_smallxi}
\begin{split}
\gamma \simeq & - \Biggr\{\frac{2}{\tau_\mathrm{s}^3} + \frac{\xi}{\tau_\mathrm{s}^2}\left(7\mathrm{Sc} + 5 + \beta_\mathrm{diff}\right) + \frac{2}{\tau_\mathrm{s}} \\ &- \xi\left[ 3\mathrm{Sc}\left(\frac{2}{9} + \beta_\mathrm{visc}\right) - 1\right]\Biggr\}  \\ &\times \left({\frac{10}{\tau_\mathrm{s}^2} + \frac{\xi}{\tau_\mathrm{s}}\left(\frac{70}{3}\mathrm{Sc} + 14 + 2\beta_\mathrm{diff}\right) + 2}\right)^{-1},
\end{split}
\end{align}
Assuming large stopping times $\tau_\mathrm{s}\gg 1$, this becomes
\begin{align}
\label{eq:gamma_overstab_approx}
    \gamma  \simeq \frac{\xi}{2}\left[3\mathrm{Sc}\left(\frac{2}{9} + \beta_\mathrm{visc}\right) - 1\right],
\end{align}
\new{which is plotted in Fig.~\ref{fig:overstability_growth_rates} in comparison to the root of the full cubic.}
For overstability, i.e. growing oscillations, the growth rate must be positive, i.e. $\gamma > 0$, or equivalently,
\begin{align}
\label{eq:viscous_overstab_criterion}
    \beta_\mathrm{visc} > \frac{\mathrm{Sc}^{-1}}{3}-\frac{2}{9}
\end{align}
which equals $1/9$ for $\mathrm{Sc} = 1$. 
\new{Since overstability is characterized by a restoring force which lacks in case of pure instability, this requirement on $\beta_\mathrm{visc}$ is opposite in direction compared to the criteria for instability discussed in the previous sections (also compare to the classical viscous overstability e.g., in \citet{Latter2006a}).}

We \new{also} note that the $1/\mathrm{Sc}$ term \new{in Eq.~\eqref{eq:viscous_overstab_criterion}} originates from advection of angular momentum carried by background shear due to the diffusive flux, i.e. the second term on the right hand side of Eq.~\eqref{eq:phi_momentum_linearized}. For $\mathrm{Sc}\gg 1$, i.e. negligible diffusion compared to viscosity, this term vanishes and we recover the classical criterion for the viscous overstability in planetary rings \citep{schmit1995}. 
This term does not allow overstability in the invicsid case.
That is, by setting $\mathrm{Sc} = 0$, the growth rate resulting from Eq.~\eqref{eq:overstab_gamma_smallxi} becomes
\begin{align}
    \gamma = - \frac{2 + \xi\tau_\mathrm{s}\left( 5 + \beta_\mathrm{diff}\right) + 2\tau_\mathrm{s}^2 + \xi\tau_\mathrm{s}^3}{10\tau_\mathrm{s} + \xi\tau_\mathrm{s}^2\left(14 + 2\beta_\mathrm{diff}\right) + 2\tau_\mathrm{s}^3},
\end{align}
which tends to $\gamma = -\xi/2$ for large stopping times $\tau_\mathrm{s} \gg 1$ \newtwo{(compare to Eq.~\eqref{eq:gamma_overstab_approx})}. 

\new{In Fig.~\ref{fig:overstability_s growth_rates_vs_tau}, we depict growth rates of the diffusive overstability obtained from Eq.~\eqref{eq:overstab_gamma_smallxi} as a function of the stopping time. The smallest scales have fastest growth rates, and, have the least stringent restrictions on stopping time. For the depicted set of parameters, $\tau_\mathrm{s} \gtrsim 10$ would lead to overstability on the smallest scales. While this requirement can be further relaxed for larger $\mathrm{Sc}$ or more positive $\beta_\mathrm{visc}$, for} $\tau_\mathrm{s} \lesssim 1$, all oscillating modes are damped.

For this reason, \new{the overstable modes discussed in this section have} little physical relevance in the context of \new{particles that generate} streaming instability turbulence, \new{where the largest available dust grains are limited at typically $\tau_\mathrm{s} \lesssim 1$ \citep[e.g.,][]{Birnstiel2016}}. We elaborate on this, \new{as well as explore alternative situations where the diffusive overstability may find applicability} in Sect.~\ref{sect:disc_overstability}.


\section{Discussion}
\label{sect:discussion}

\begin{figure}
    \centering
    \includegraphics[width = \linewidth]{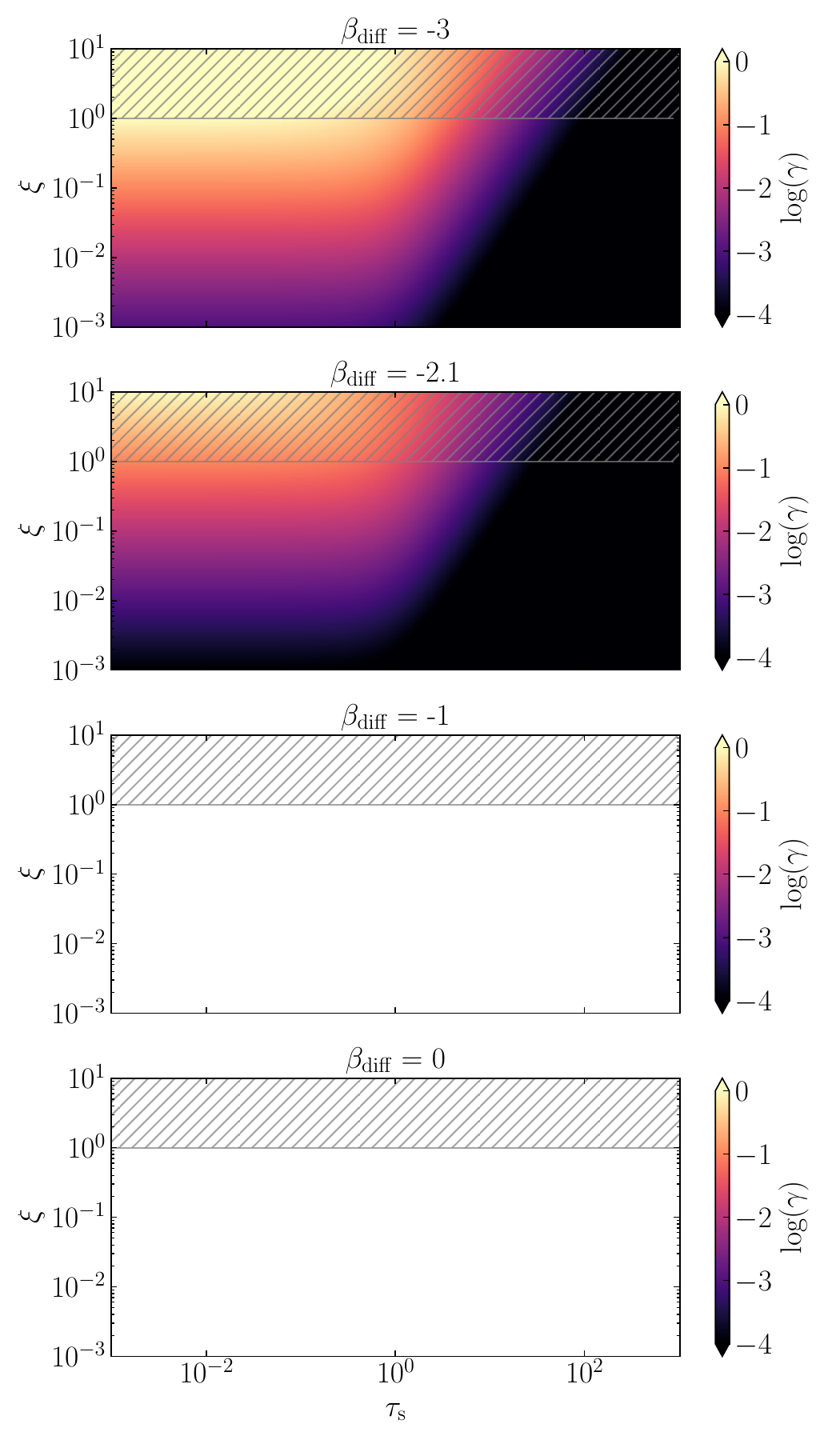}
    \caption{Growth rates for inviscid ($\mathrm{Sc} =0$) diffusive instability. The diffusion slope $\beta_\mathrm{diff}$ \new{increases} from top to bottom. The diffusive instability driven by diffusion-dependent pressure is active for all modes $\xi$ if $\beta_\mathrm{diff} < -2$ (see  Eq.~\eqref{eq:inviscid_diffusive_instability_criterion}), which is the case in the upper two panels for small stopping times. For large stopping times, growth rates become uninterestingly small. $\mathrm{Sc} = 0$, does not allow growth of oscillatory modes. 
    }
    \label{fig:all_growth_rates_inviscid}
\end{figure}

\begin{figure}
    \centering
    \includegraphics[width = \linewidth]{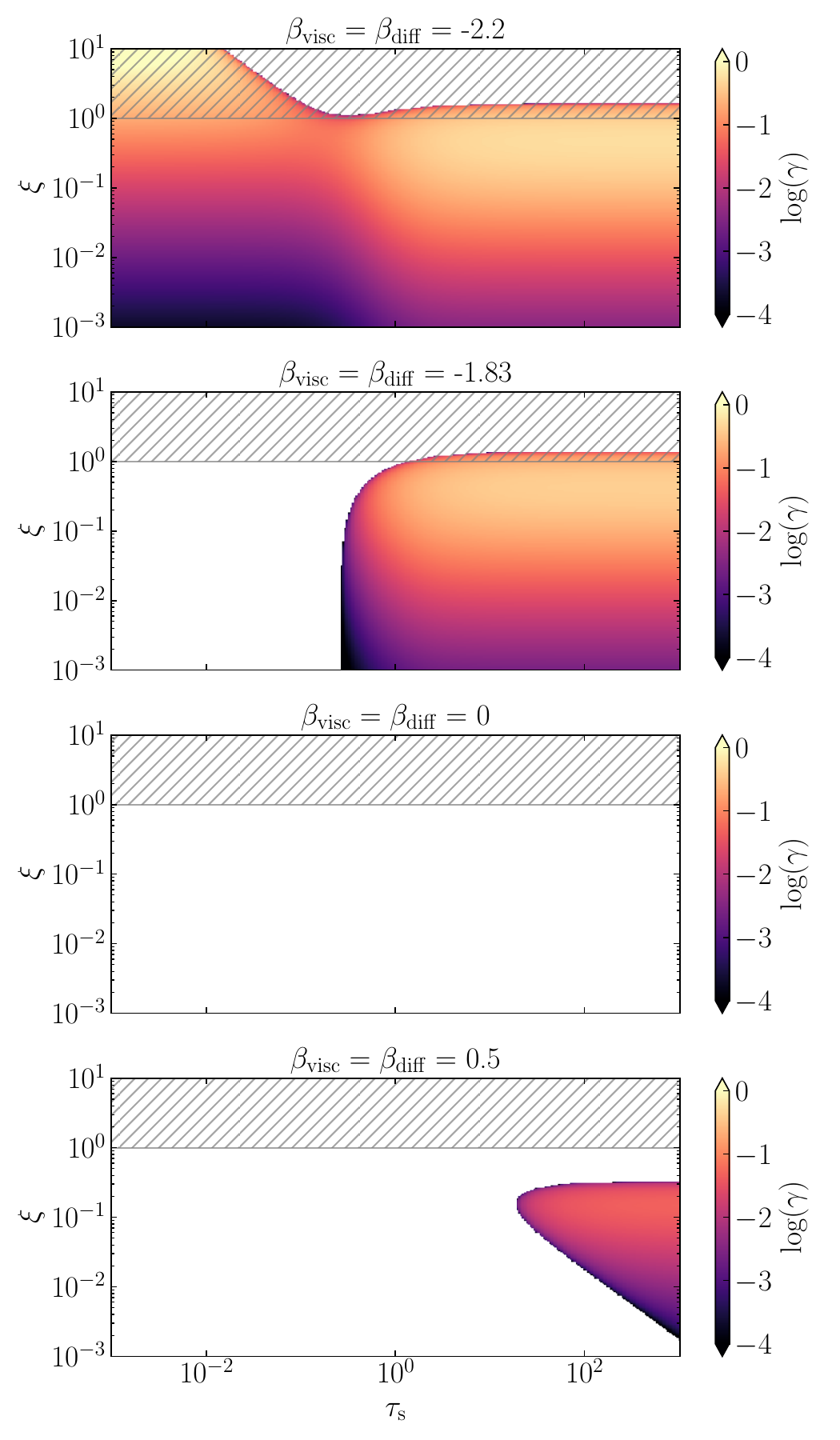}
    \caption{Like Fig.~\ref{fig:all_growth_rates_inviscid} but for the viscous-diffusive instability and (modified) viscous overstability with $\mathrm{Sc} =1$. The top panel shows combination of diffusive instability driven by diffusion-dependent pressure and viscous instability driven by the negative viscosity slope for small and large stopping times respectively (compare to Fig.~\ref{fig:Sc=1_diff_inst_xi0.1}). In the panel second from the top, \newtwo{Eq.~\eqref{eq:Sc1_pureinstcrit} remains satisfied, while} Eq.~\eqref{eq:inviscid_diffusive_instability_criterion} is not, \newtwo{so instability is driven by the viscosity slope alone and thus restricted to larger $\tau_\mathrm{s}$}. In the third panel from the top, neither instability is active. When Eq.~\eqref{eq:viscous_overstab_criterion} is satisfied (for $\mathrm{Sc} = 1$ when $\beta_\mathrm{visc}$ \new{exceeds $+1/9$}, the (modified) viscous overstability driven by the viscosity slope can be active for large stopping times. 
    }
    \label{fig:all_growth_rates_viscous}
\end{figure}

In the previous sections, we showed that a dust fluid in protoplanetary disks, which is governed by Eqs.~\eqref{eq:continuity_first} ---\eqref{eq:momentum_first_phi} can be unstable (Sect.~\ref{sect:instability}) and over-stable (Sect.~\ref{sect:overstability}) when an otherwise constant background state is linearly perturbed, depending on the stopping time of particles and the steepness of the diffusion and viscosity slopes with respect to the dust surface mass density. Here, we discuss and contextualize our findings.

\subsection{Physical picture}
\label{sect:physicalpicture}

First, we reiterate the physical mechanisms that drive the newly found instabilities. Figures~\ref{fig:all_growth_rates_inviscid} and \ref{fig:all_growth_rates_viscous} show the growth rates of the full system including the inviscid ($\mathrm{Sc} = 0$) and viscous  ($\mathrm{Sc} = 1$). Guided by the depicted parameter space we can broadly assign the unstable regions into three categories
\begin{enumerate}
    \item \textit{Diffusive instability driven by diffusion-dependent pressure}: For small stopping times, the utilized dust pressure prescription (with velocity dispersion $c_\mathrm{d}^2\propto D \propto \Sigma^{\beta_\mathrm{diff}}$) can lead to linear instability if the diffusion slope $\beta_\mathrm{diff}$ is sufficiently negative (see \newtwo{Eqs. ~\eqref{eq:inviscid_diffusive_instability_criterion} and \eqref{eq:Sc1_pureinstcrit}}), and drives particles towards density maxima. The instability is only damped on small scales if viscosity is nonzero, and has fastest growth rates on the smallest unstable scales. This instability requires $\tau_\mathrm{s} \lesssim 1$, $\beta_\mathrm{diff} < -2$, and if $\mathrm{Sc} \gtrsim 1$ also $\xi \lesssim 1$, which for a fiducial $\delta \sim 10^{-5}$ corresponds to wavelengths of $\lambda \gtrsim 0.01 H$. Associated growth rates are shown in the top two panels of Fig.~\ref{fig:all_growth_rates_inviscid} as well the left side of the top panel of Fig.~\ref{fig:all_growth_rates_viscous} for small stopping times.
    
    \item \textit{Diffusive instability driven by viscosity slope}: For nonzero viscosity, the viscosity term dominates over the pressure term for sufficiently large stopping times, leading to a version of the viscous instability \citep[][]{ward1981, LinBodenheimer1981}. This instability requires $\tau_\mathrm{s} \gtrsim 1$, $\mathrm{Sc} \gtrsim 1$, $\xi \lesssim 1$,  $\beta_\mathrm{visc} \lesssim -1$, and is seen in the top two panels of Fig.~\ref{fig:all_growth_rates_viscous} for large stopping times.

    
    \item \textit{Diffusive overstability driven by viscosity slope}: Analogously to the classical viscous overstability \citep{schmit1995}, the repellent term that amplifies oscillations is provided by a viscosity slope. This overstability requires $\tau_\mathrm{s} \gtrsim 10$, $\mathrm{Sc} \gtrsim 1$, $\xi \lesssim 1$, and $\beta_\mathrm{visc} \gtrsim +1/9$ (depending on the value of $\mathrm{Sc}$). The associated growth rates are shown in the bottom panel of Fig.~\ref{fig:all_growth_rates_viscous}.
\end{enumerate}

The linear theory describing the diffusive instabilities and overstabilities presented in
this work is largely similar to that of the classical viscous instability \citep{ward1981, LinBodenheimer1981} and axisymmetric viscous overstability \citep{schmit1995} in planetary rings, at least under the neglect of self-gravity and thermal effects, and given the appropriate limits in our model, i.e. $\tau_\mathrm{s} \gg 1$, $\mathrm{Sc} \gg 1$. 
It should be noted though that the physical origin of viscosity and pressure in planetary rings lies in mutual particle collisions, in contrast to the situation depicted in this work, where it results from self-generated turbulence, that is external insofar as the model is concerned. Also, a `thin disk' version of the viscous overstability (i.e. on large radial length scales $\gg H_{g}$), generated by a constant kinematic shear viscosity, can in principle also exist in gaseous proto-planetary disks \citep{Latter2006a}.

In addition, we mention the dust-driven viscous ring-instability pioneered by \citet{Dullemond2018} as well as the related instability in \citet{Johansen2011}, both of which, although operating on larger scales, are similar in spirit to this paper's diffusive instabilities. They consider a disk where the gas viscosity is set by turbulence generated from magnetorotational instability \citep{Balbus1991}, which weakens with increasing dust density through the ionization fraction. It is the viscous gas disk itself that can now be unstable to linear perturbations: an increase in gas density attracts dust grains as they tend to drift towards gas pressure maxima \citep[e.g.][]{Sano2000}, turbulence weakens and the gas viscosity drops, thereby increasing the gas density further, which attracts more dust, and so on. 
In our model, we only consider the dust fluid but the decrease in its viscosity, diffusivity, and dust pressure as the dust surface density increases is likewise physically motivated by dust feedback lowering the local diffusive properties of the turbulence generated on small-scales.

Indeed, while we motivate our model with streaming instability turbulence and associated measurements of diffusion slopes \citep{Schreiber2018, Gerbig2023}, it may as well be applicable to other sources of turbulence. For example, the azimuthal streaming instability discovered by \citet[][]{Hsu2022} has likewise been observed to evolve into filaments once linear modes are saturated. Moreover, if one considers pure gas instabilities as sources of turbulence, the vertical shear instability \citep[][]{Urpin1998, Nelson2013, Lin2015, Barker2015, Pfeil2021} or the convective overstability \citep{Klahr2014, Lyra2014, Latter2016} may generate environments suitable for secondary diffusive instabilities. The dependence of diffusivity and viscosity on disk parameters would need to be clarified with detailed simulations, but to zeroth order, one expects a drop in turbulence with increasing dust-loading, similar to that for the magnetorotational instability discussed above, because dust feedback tends to stabilize the vertical shear instability \citep{Lin2019,lehmann2022}, and linear convective overstability \citep{Lehmann2023}.

\subsection{Filament formation through diffusive instability}

\begin{figure}
    \centering
    \includegraphics[width = \linewidth]{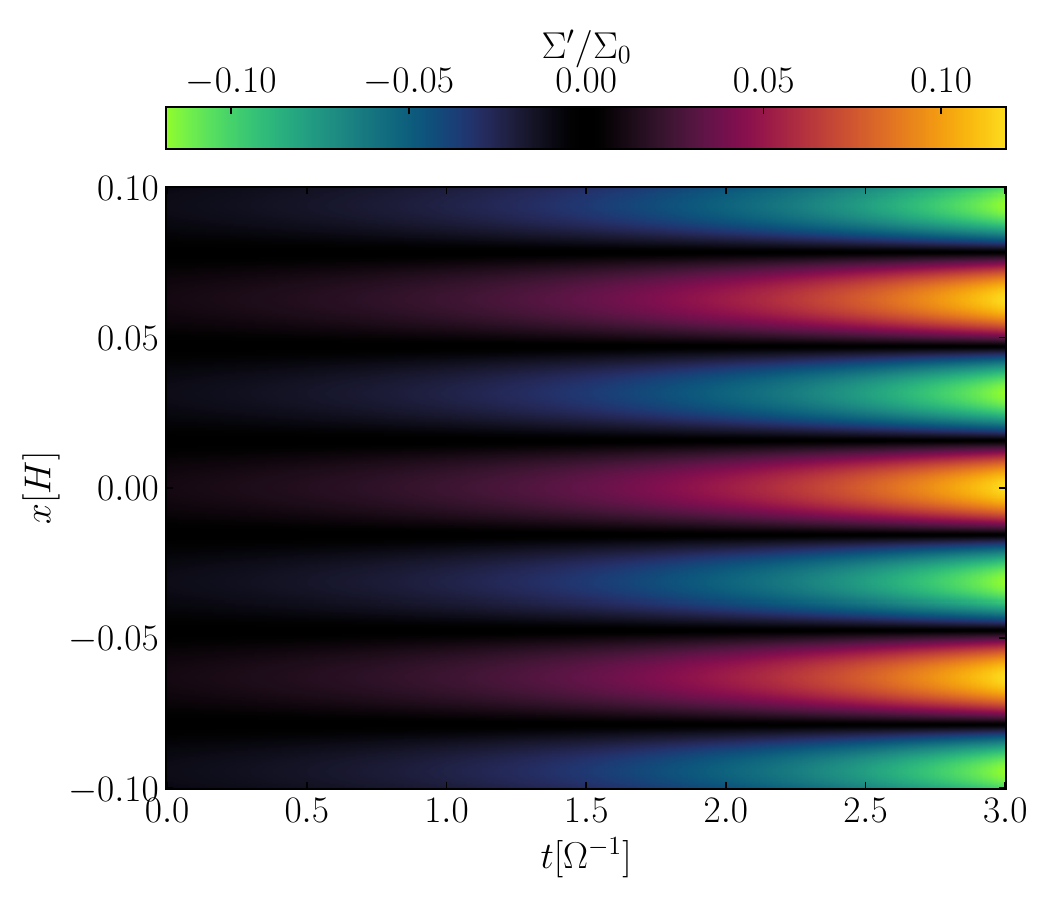}
    \caption{Filament formation shown in a space time diagram of a linear perturbation subject to diffusive instability driven by pressure. For this plot, we chose $\mathrm{Sc} = 0$, $\tau_\mathrm{s} = 0.1$, $\beta_\mathrm{diff} = -3$. We also chose a value of $\delta = 10^{-4}$, and a fast growing mode of $\xi = 0.1$, this corresponds to a physical wave number of $k = 100/H$. The eigenvector is scaled with $\hat{\Sigma} = 0.01 \Sigma_0$. 
    } 
    \label{fig:inst_space_time}
\end{figure}

\begin{figure}
    \centering
    \includegraphics[width = \linewidth]{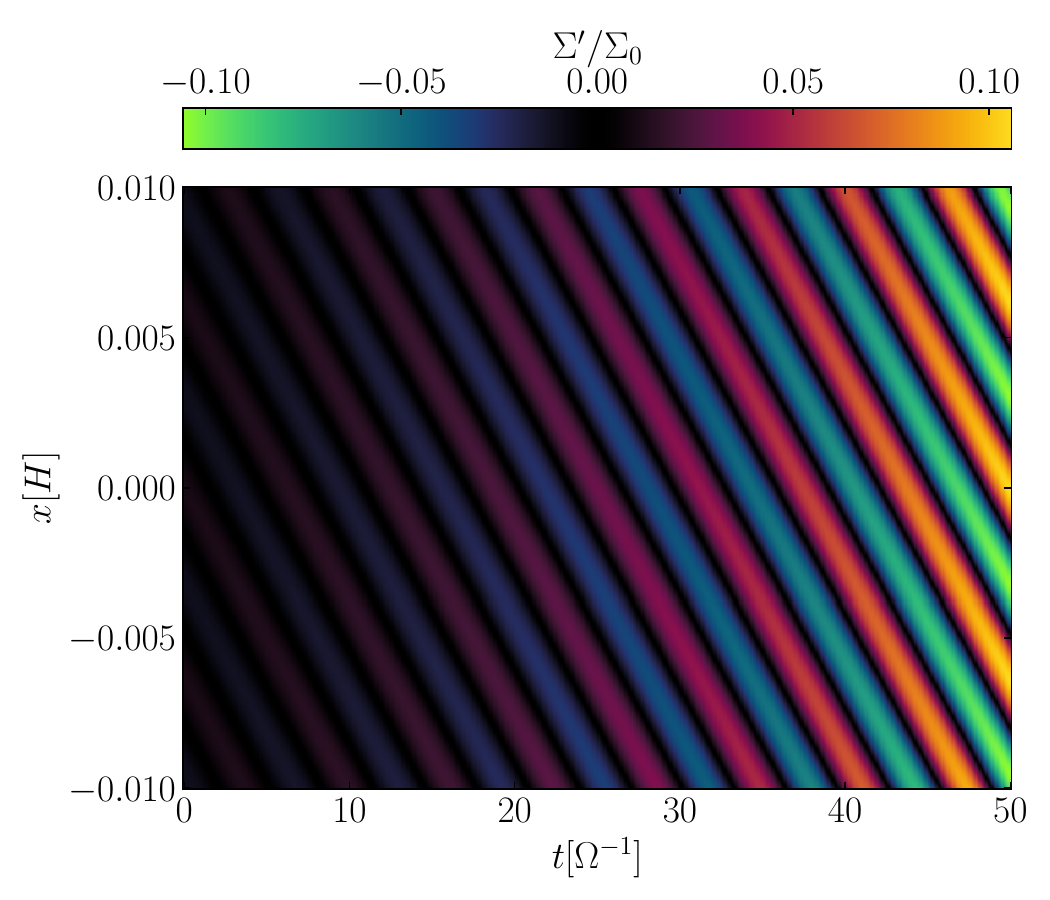}
    \caption{Like Fig.~\ref{fig:inst_space_time}, but for an oscillatory mode. The linear perturbation is subject to diffusive overstability powered by the viscosity slope. We chose, $\mathrm{Sc} = 1$, $\tau_\mathrm{s} = 1000$, $\beta_\mathrm{diff} = \beta_\mathrm{visc} = 1/2$, and $\hat{\Sigma} = 0.01 \Sigma_0$. We took $\delta = 10^{-7}$, at which a mode of $\xi = 0.1$ corresponds to a physical wave number of $k = 1000/H$. Note, that the shearing sheet is symmetric in $x$, so waves traveling towards positive $x$ are equally valid.}
    \label{fig:over_space_time}
\end{figure}

\begin{figure}
    \centering
    \includegraphics[width = \linewidth]{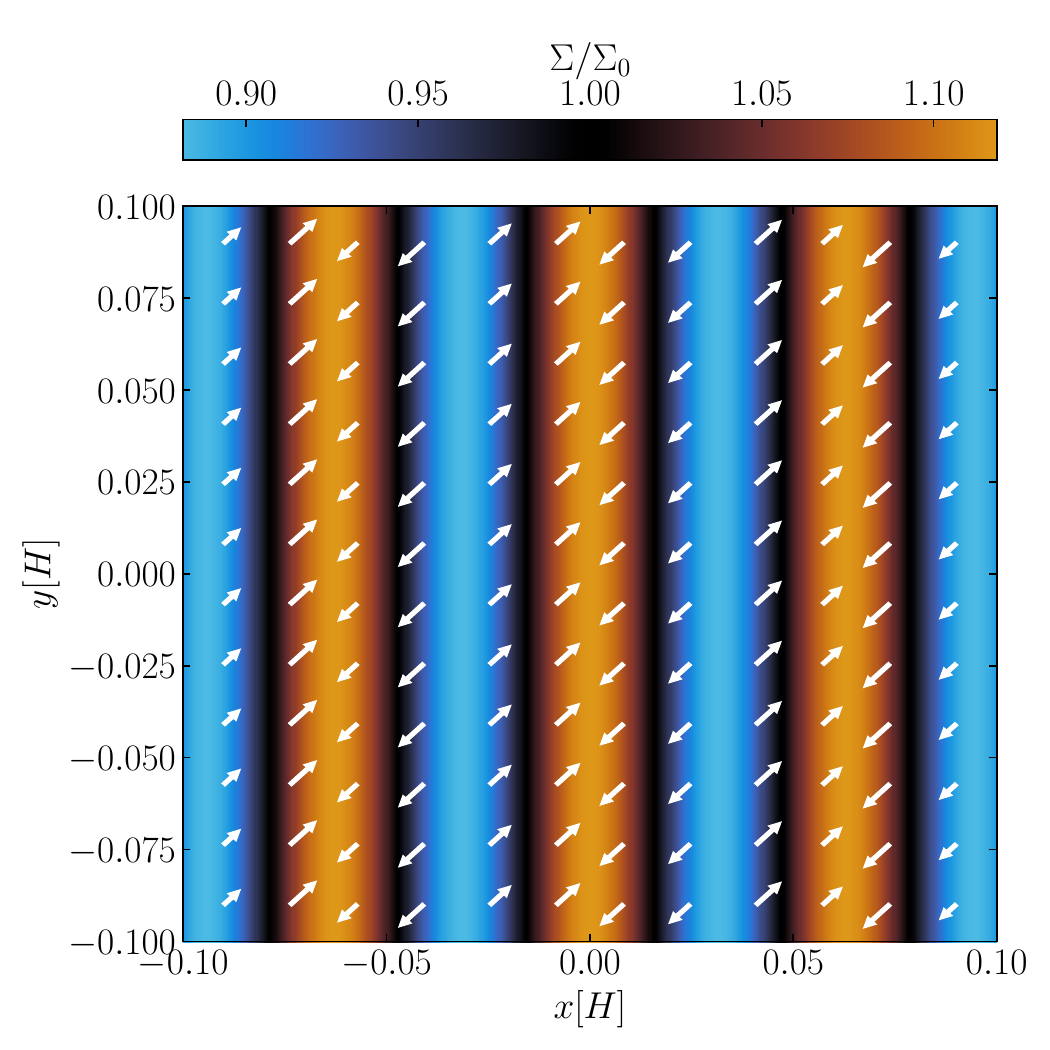}
    \caption{Diffusive instability driven by diffusion-dependent pressure with same parameters as in Fig.~\ref{fig:inst_space_time}. Shown is a map in $x-y$ space of the particle surface density after a time of $3\Omega^{-1}$, with super-imposed re-scaled velocity vectors $(v_x^\prime, v_y^\prime)$. The fluid motion is towards density maxima, thus amplifying the perturbation and producing filaments. We show the $y$-coordinate for better visualization, even though the model itself is axisymmetric.}
    \label{fig:vel_orientation}
\end{figure}

Based on the analytic findings in Sects.~\ref{sect:instability} and \ref{sect:overstability}, we hypothesize, that the diffusive instability, driven by a sufficiently negative density dependence of the dust pressure, physically motivated by dust loading reducing diffusivity, may act to amplify density perturbations. These perturbations then could saturate to become marginally stable filaments, as seen in many past simulations \citep[e.g.,][]{Johansen2007, Johansen2009, Carrera2015, Klahr2016, Yang2018, Li2018, Schreiber2018, Sekiya2018, Gerbig2020, Flock2021, Li2021, Hsu2022, Gerbig2023}. 

We visualize the growth of the linear perturbations into such filament-like overdensities in a space - time diagram of the perturbed density $\Sigma^\prime$ following Eq.~\eqref{eq:perturbuation_def}. Fig.~\ref{fig:inst_space_time} shows a perturbation subject to diffusive instability. The chosen set of parameters produces overdensities with a radial spacing of about $0.02H$, which is consistent with the first emergent filaments found in streaming instability simulations \citep[e.g.,][]{Li2021}. 

Indeed, the diffusive instability has the fastest growth rates on the smallest scale not limited by viscosity $\xi_\mathrm{max}$, which is of order $\xi_\mathrm{max} \sim 1$. The corresponding fastest growing mode is thus expected to be around
\begin{align}
\label{eq:diff_inst_fgm}
    \frac{\lambda_\mathrm{fgm}}{H} = \frac{2\pi}{K} = 2\pi\sqrt{\frac{\delta}{\xi_\mathrm{max}}} \approx 2\pi\sqrt{\delta}
\end{align}
For $\delta \sim 10^{-5}$ \citep[e.g.][]{Schreiber2018, Klahr2020, Gerbig2023} this would correspond to a consistent value of $\lambda_\mathrm{fgm} \sim 0.02 H$ \citep[e.g.,][]{Li2021}. Note that, the filament separation in simulations has been found to depend on external gas pressure slope, stopping time, and dust abundance \citep[e.g.][]{Schreiber2018, Gerbig2020, Li2021}. If these properties influence streaming instability turbulence \citep[e.g.,][]{Johansen2007}, they are expected to map onto diffusivity and ultimately on the fastest growing mode in Eq.~\eqref{eq:diff_inst_fgm}. On the other hand, the commonly used fiducial value for the filament feeding zone of $0.2 H$ \citep{Yang2014}, cannot be directly compared to the scales of interest, as it already involves post-formation nonlinear dynamics, such as mergers and breakups.

Fig.~\ref{fig:vel_orientation} visualizes the streaming motion that arises from the diffusive instability. Dust is moving towards density maxima which act as particle traps, in the process amplifying the perturbation. 

For comparison, Fig.~\ref{fig:over_space_time} shows traveling waves driven by the diffusive overstability. The shearing sheet is symmetric in $x$, so waves traveling towards positive $x$ correspond to the complex conjugate of the depicted solution and are equally valid. Note, that the radial bulk velocity is entirely caused by to the overstability. If one were to include the background pressure gradient, the flow would pick up a to-first-order, constant background drift, which does not affect stability since we can shift into the co-moving frame.

\subsection{Diffusive instability in the context of past numerical simulations}

The aim of this study is to present a simple linear model for the emergence of the first filaments out of streaming instability turbulence,  as seen in a number of simulations including  \citet{Schreiber2018, Gerbig2020, Li2021, Gerbig2023}.
Within our model, the required linear growth rates and corresponding required stopping times that are expected to result in filament formation depend on the values of the diffusion and viscosity slopes $\beta_\mathrm{diff}$ and $\beta_\mathrm{visc}$, respectively. Therefore, we aim to contextualize our findings with existing numerical studies and measurements of the aforementioned slopes. 
As such, our presented theory can most readily be compared to the numerical results presented in \citet[][]{Schreiber2018}, who performed two-dimensional, non-axisymmetric, non-selfgravitating shearing sheet simulations of the streaming instability. By conducting a number of simulations at different dust-to-gas ratios and measuring the average particle diffusivity in each simulation, they were able to obtain the slope of diffusion with respect to the dust surface mass density, resulting in values \new{$-2 \lesssim \beta_\mathrm{diff} \lesssim -1$}, where the exact value depended on box size and particle stopping time used in their simulations. 

While some of the simulations of \citet{Schreiber2018} revealed the emergence of  filaments with particle concentrations significantly enhanced relative to the ambient background, most did not. In the context of the linear diffusive instability discovered in our work, which requires $\beta_\mathrm{diff} < -2$ (see Fig.~\ref{fig:Sc=1_diff_inst_xi0.1}) for instability, the simulations in \citet{Schreiber2018} are hence expected to be marginally stable. More recently, \citet{Gerbig2023} measured the diffusion slope within a single vertically stratified  shearing box simulation of the streaming instability. They also found $\beta_\mathrm{diff} \sim -2$. As this value was obtained \emph{after} the formation of filaments in their simulations, this is also consistent with their simulations being marginally unstable with respect to diffusive instability, provided the saturation of filament growth results in such a marginally-stable state.

\subsection{Connection to planetesimal formation}
\label{sect:plsformation}

\begin{figure}
    \centering
    \includegraphics[width = \linewidth]{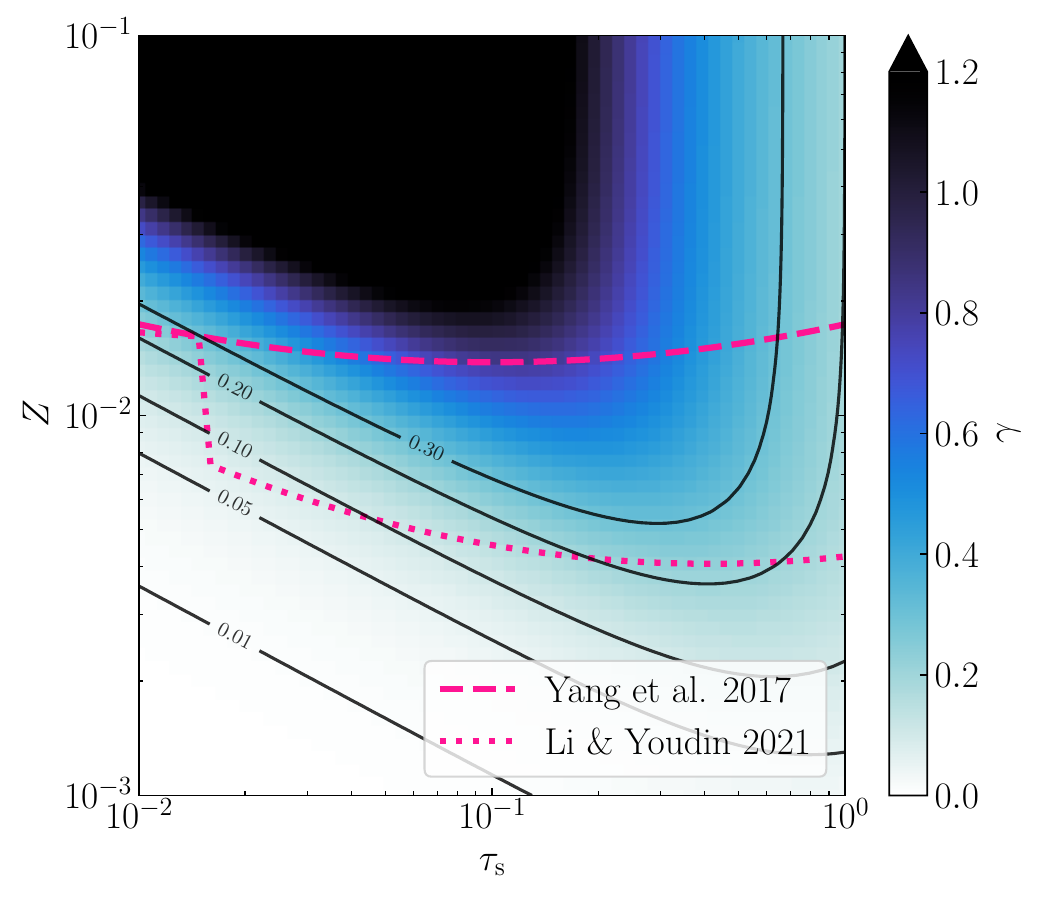}
    \caption{Growth rates of diffusive instabilty driven by diffusion-dependent pressure term, depending on metallicity $Z$ and stopping time $\tau_\mathrm{s}$, compared to clumping thresholds in streaming instability simulations by \citet{Yang2017} and \citet{Li2021}. We associate a given metallicity $Z$ with the fastest growing, allowed mode $\xi$ of our model via the recipe discussed in Sect.~\ref{sect:plsformation} assuming $\epsilon = 1$, $\lambda_\mathrm{crit} = 0.01 H$. We also chose $\beta_\mathrm{diff} = \beta_\mathrm{visc} = -2.2$ and $\mathrm{Sc} =0$. 
    }
    \label{fig:thresholds_inviscid}
\end{figure}

\begin{figure}
    \centering
    \includegraphics[width = \linewidth]{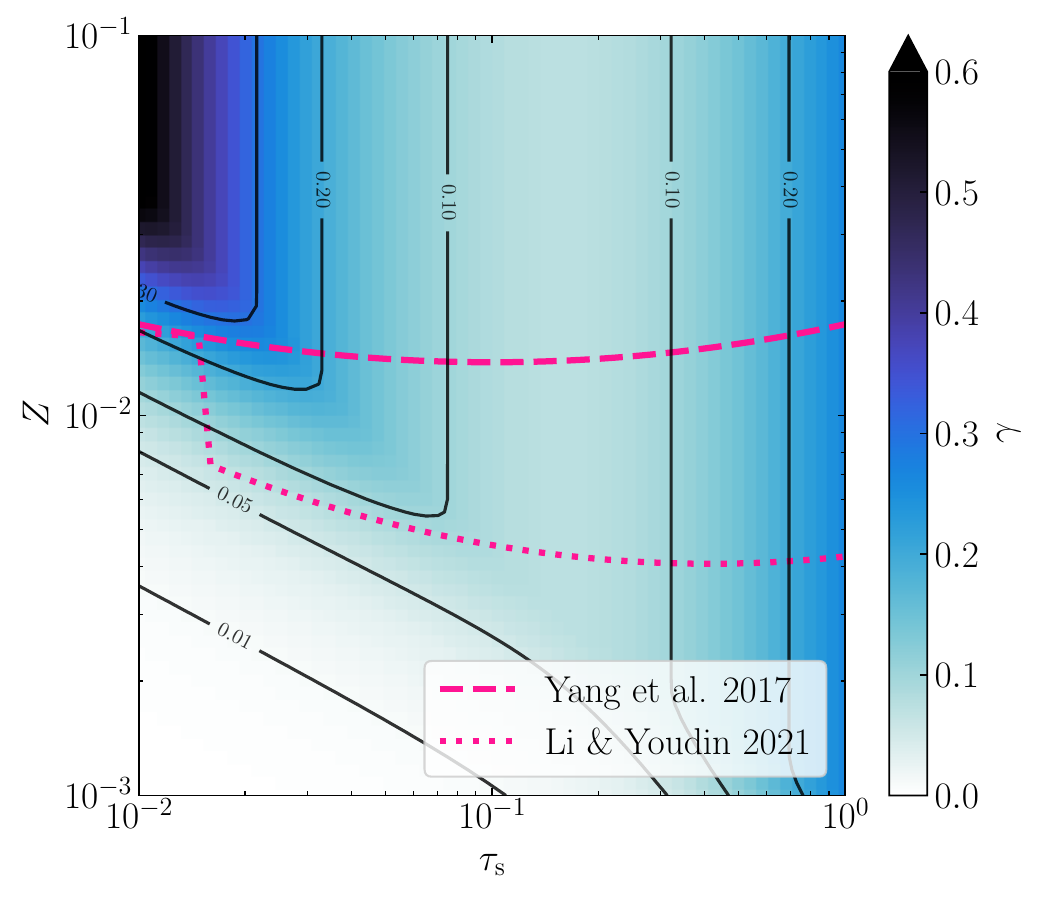}
    \caption{Like Fig.~\ref{fig:thresholds_inviscid} but now for $\mathrm{Sc} = 1$, which allows for instability driven by the viscosity slope at large stopping times in addition to the instability driven by the diffusion-dependent pressure term. Other parameter choices are identical to those in Fig.~\ref{fig:thresholds_inviscid}.}
    \label{fig:thresholds_viscous}
\end{figure}

Within the streaming instability paradigm of planetesimal formation, filaments are often thought to be a necessary precursor for planetesimal formation, as they provide the necessary conditions for subsequent self-gravitational fragmentation. It is therefore of interest to compare the parameter space for active diffusive instability within our model to that determined in numerical simulations of planetesimal formation, specifically clumping thresholds in streaming instability simulations. For this purpose, we can relate metallicity $Z$ to the dust-to-gas ratio $\epsilon = \rho_\mathrm{p}/\rho_\mathrm{g}$ and diffusivity via \citep[e.g.,][]{Lin2021} 
\begin{align}
\label{eq:Zfromscaleheight}
    Z = \frac{\Sigma}{\Sigma_\mathrm{g}} \simeq \epsilon \frac{H_\mathrm{d}}{H}  \simeq \epsilon \sqrt{\frac{\delta}{\delta + \tau_\mathrm{s}}},
\end{align}
where $H_\mathrm{d}$ is the vertical scale height of dust.

Assuming $\tau_\mathrm{s}\gg \delta$, we have $\delta \sim \tau_\mathrm{s}(Z/\epsilon)^2$. Since our model, and therefore also the diffusive instability mechanism only implicitly depends on $\delta$ via $\xi$, a change in diffusivity due to metallicity increase (decrease) can be compensated by a decrease (increase) in wavenumber $K$ towards larger (smaller) radial length scales. For example, the typically fast growing mode of $\xi \sim 1$, would correspond to wave numbers of $K \sim \epsilon/(Z\sqrt{\tau_\mathrm{s}})$.

We proceed by imposing a minimum scale of $\lambda_\mathrm{crit} \sim 0.01 H$, below which the model does not apply (cf. Sect.~\ref{sect:governing_eqs}), and use this to restrict the maximum allowed $\xi$, given some value of $Z$. Specifically, we have
\begin{align}
\label{eq:ximaxofZtau}
    \xi_\mathrm{max}(Z,\tau_\mathrm{s}) =  \frac{4\pi^2\delta(Z,\tau_\mathrm{s})}{(\lambda_\mathrm{crit}/H)^2} \simeq  \left(\frac{2\pi Z/\epsilon}{\lambda_\mathrm{crit}/H}\right)^2 \tau_\mathrm{s}.
\end{align}
We then assume that the fastest growing mode admitted by our model is given by $\xi(Z, \tau_\mathrm{s}) = \min(\xi_\mathrm{fgm}, \xi_\mathrm{max})$, where $\xi_\mathrm{fgm}$ is the mathematically fastest growing mode, which \new{is obtained numerically from the full dispersion relation in Eq.~\eqref{eq:disprel_sigma}. $\xi_\mathrm{fgm}$ is} typically of order unity (see Fig.~\ref{fig:inviscid_diff_inst_growthrates}), but may be larger, \new{either in the absence of viscosity or for very small stopping times (see Figs.~\ref{fig:inviscid_diff_inst_growthrates} and ~\ref{fig:visc_diff_inst_growthrates} respectively)}; or smaller, if the viscosity slope is only marginally more negative than the critically required slope (see top panel in Fig.~\ref{fig:all_growth_rates_viscous}). 

Figures~\ref{fig:thresholds_inviscid} and \ref{fig:thresholds_viscous} show the growth rates associated with this preferred mode assuming $\epsilon = 1$, and $\lambda_\mathrm{crit} = 0.01H$. We also take $\beta_\mathrm{diff} = \beta_\mathrm{visc} = -2.2$ in both Figures. Fig.~\ref{fig:thresholds_inviscid} shows the inviscid case with $\mathrm{Sc} =0$, where only the diffusive instability driven by diffusion-dependent pressure is possible given the set of parameters, and Fig.~\ref{fig:thresholds_viscous} shows growth rates associated with the same set of parameters, except that now $\mathrm{Sc} = 1$. Over-plotted are the clumping thresholds obtained by \citet{Yang2017} and \citet{Li2021} (also see \citet{Carrera2015} for another version), below which, streaming instability clumping is unlikely to occur. 

For either model, the entire parameter space, probed by the studies \citet{Yang2017} and \citet{Li2021} would be subject to diffusive instabilities (although with differences in expected growth rates), rendering the models consistent with the hypothesis that planetesimal formation is triggered by the emergence of filaments induced by diffusive instability.

We note, that the models displayed in Figs.~\ref{fig:thresholds_inviscid} and \ref{fig:thresholds_viscous} contain a number of parameters ($\lambda_\mathrm{crit}, \epsilon, \beta_\mathrm{diff}, \beta_\mathrm{visc}, \mathrm{Sc}$), which we chose relatively freely \new{and assumed to be independent of $\tau_\mathrm{s}$} in order to get a prototype idea if there may be a connection to planetesimal formation. 
More detailed calculations, in concert with additional numerical constraints on diffusion and viscosity slopes, are required to further assess diffusion instability's role in planetesimal formation.


\subsection{Instability and overstability at large stopping times?}
\label{sect:disc_overstability}

The local axisymmetric viscous overstability of a thin astrophysical disk has extensively been studied in the context of Saturn's dense rings, employing hydrodynamic models \citep{schmit1995,schmit1999,spahn2000a,Schmidt2001,schmidt2003,Latter2009,latter2010,lehmann2017,lehmann2019},  kinetic models \citep{latter2006b,latter2008}, and N-body simulations \citep{salo2001,latter2013,ballouz2017,lehmann2017,mondino2023}. Based on results from hydrodynamic and N-body simulations, overstability in Saturn's rings typically saturates in the form of nonlinear traveling wave trains that could in principle carry appreciable amounts of angular momentum. 
Wave trains are often interspersed by defect structures, which may act as sources or sinks of the former. 
Indeed, viscous overstability is the most promising mechanism to explain the occurrence of periodic fine structures on a $\sim100$m scale in parts of Saturn's A and B rings, that has directly been observed \citep{thomson2007,colwell2007,hedman2014a}. It is thus of interest to explore the conditions under which we expect the diffusive overstability to operate in over-dense particle layers in protoplanetary disks.

Since this requires $\tau_\mathrm{s} \gtrsim 1$, we preface further discussion, by reiterating that the hydrodynamical model is strictly not applicable for large stopping times, and instead a kinetic model should be used (see Sect.~\ref{sect:applicability}). 

As outlined in Sect.~\ref{sect:physicalpicture}, in addition to the diffusive instability driven by the pressure term, there are two types of instabilities that can arise for large stopping times, i.e. instability and overstability driven by the viscosity slope with their respective specific requirements on on $\beta_\mathrm{visc}$. 

The questions are (a) to what extent the invisicd, hydrodynamic model can still appropriately describe the system for $\tau_{s} \gtrsim 1$, and whether or not particles with sufficiently large stopping times (b) can exist and (c) are \new{in their turbulent behavior} still well \new{characterized} by the diffusive flux model that shapes the continuity equation, the pressure model and the angular momentum conserving terms in the momentum equations.

Indeed, in the classical picture of particle growth in protoplanetary disks, stopping times are limited at around $\tau_\mathrm{s} \sim 1$ \new{\citep[e.g.,][]{Birnstiel2012, Birnstiel2016}}, and as such, individual particles that qualify for diffusive overstability would already be considered planetesimal-sized objects. Another possible pathway of getting objects with large stopping times was suggested by \citet{Johansen2007}. \new{In an effort to explain unexpected drift rates of particle clumps found in their streaming instability simulations}, they hypothesize that a clump may collectively have an increased stopping time relative to the individual grains, due to shielding each other from the gas stream and providing an order-of-magnitude scaling 
\begin{align}
    \tau_\mathrm{s}^\mathrm{eff} \sim \epsilon \frac{R_\mathrm{clump}}{\eta r} \frac{\eta r \Omega}{\Delta v},
\end{align}
with $R_\mathrm{clump}$ being the clump's radius, and $\Delta v$ its velocity relative to the gas. 
\new{While such clumps are far too few in number density to be appropriately modeled by a fluid approach, we adapt this hypothesized collective shielding effect to our situation.}

\new{Specifically,} applying \citeauthor{Johansen2007}'s argument to the quasi-steady, turbulent dust layer that we envision as the equilibrium, we set $R_\mathrm{clump} = H_\mathrm{d}$, the dust scale height. Then, using $H_\mathrm{d} = ZH/\epsilon$ from Eq.~\eqref{eq:Zfromscaleheight}, we find $\tau_\mathrm{s}^\mathrm{eff}\sim Z/\Pi$, where $\Pi \equiv \eta/(H/r)$ is the reduced radial pressure gradient parameter, again taking $\Delta v\sim\eta r \Omega$. The diffusive overstability's requirements of $\tau_\mathrm{s}^\mathrm{eff}\gtrsim 1$ translates to $Z\gtrsim \Pi$, and the same applies for the  diffusive instability driven by viscosity slope. Coincidentally, \cite{bai10} find that clumping via the streaming instability becomes easier with smaller $\Pi$. Similarly, at fixed $\tau_\mathrm{s}$ (the stopping time for individual grains), \cite{Sekiya2018} find filament formation in their streaming instability simulations if  $\sqrt{2\pi}Z/\Pi\gtrsim 1$. We can interpret these results within our model as filaments only form if the effective stopping time, realized through $Z/\Pi$, exceeds unity in order to trigger the diffusive overstability (or the diffusive instability driven by viscosity slope).

\new{The streaming instability formally still operates for $\tau_\mathrm{s} \gtrsim 1$ as growth rates only decrease slowly with stopping time \citep{Pan2020}. However, due to the lack of numerical constraints on diffusivity in this regime, it is unclear to what extent instabilities and overstabilities would develop on scales within our model's validity; even if diffusion and viscosity slopes were to remain unchanged.}
\new{Since,} in order to achieve instability on relevant scales, diffusivities cannot be arbitrarily small, \new{we consider the possibility of diffusion to not be self-generated.} Instead, diffusion \new{may} stem directly from a turbulent gas, the diffusivity and viscosity of which we denote as $\delta_\mathrm{g}$ and $\alpha_\mathrm{g}$ respectively.

In the preceding sections, we were exclusively concerned with particle diffusivity $\delta$, which we treated as wholly independent from $\tau_\mathrm{s}$. While this is mathematically self-consistent, it does not necessarily reflect physical conditions. Indeed, an increased particle response time to turbulence diminishes the diffusion experienced by the particles as \citep{Youdin2011}
\begin{align}
\label{eq:diffusion_dependence}
    \delta = \frac{1+\tau_\mathrm{s} + 4\tau_\mathrm{s}^2}{(1 +\tau_\mathrm{s}^2)^2}\delta_\mathrm{g}, 
\end{align}
which reduces to $\delta \sim \delta_\mathrm{g}$ for small stopping times, but becomes $\delta \sim \delta_\mathrm{g}/\tau_\mathrm{s}^2$ large stopping times. That is, large grains feel a much reduced turbulence than that in the gas.

Consider, for example, gas with a fiducial diffusivity of $\delta_\mathrm{g} \sim \alpha_\mathrm{g} \sim 10^{-3}$, which is a typical value in numerical simulations of vertical shear instability or magnetorotational instability \citep[e.g.,][]{Flock2017}. For $\tau_\mathrm{s} \sim 10$, this would lead to $\delta\sim 10^{-5}$, that is comparable to the diffusivities generated by streaming instability with smaller particles. The fastest growing scale per Eq.~\eqref{eq:diff_inst_fgm} remains unchanged at $\sim 0.02 H$. Indeed, since the diffusive instabilities discussed in this paper depend on $\xi = \delta K^2$ only, any decrease in diffusivity due to particle response to turbulence would only shift the fastest growing mode down to smaller scales, but not prohibit the mechanism itself from operating. 

We thus argue that the large stopping time instability and overstabilities may find applicability, if a big enough collection of large stopping time particles are available in protoplanetary disks, or if the dust layer has a collective stopping time that exceeds unity. Note, that this estimation ignored the effect of the particle layer on gas turbulence, i.e. $\delta_\mathrm{g}$ itself. For example, turbulence driven by the vertical shear instability is damped by dust feedback, even when the dust-to-gas ratio is less than unity \citep[e.g.,][]{Lin2019}.


\subsection{Caveats and additional considerations}

Our vertically averaged model neglects vertical motions. Therefore, inertial waves are discarded, such that the classical, axisymmetric streaming instability is not captured by our model \citep{Squire2018}, even if we were to include the gas equations with a radial background pressure gradient. While we attribute the underlying turbulence, characterized by diffusion and viscosity of the dust fluid, to the streaming instability (or an equivalent mechanism), filament formation in our model is not a direct result of this underlying, small-scale instability. Instead, it originates from one or more intrinsic large-scale instabilities of the mid-plane dust layer, described by our model. \new{In order to neglect gas perturbations, our model is time-averaged over one turbulent correlation time. As a result, linear modes with frequency smaller than one inverse correlation time should be considered with caution.} Future work should include the dynamical equations describing the gas, as well as the vertical dimension, to investigate filament formation via streaming-type instabilities with variable viscosity and diffusivity in a more rigorous manner. 

In our model, the diffusion and viscosity slope is assumed to depend on dust surface density only, primarily because of the available numerical constraints from \citet{Schreiber2018, Gerbig2023} and due to the analogy to isothermal hydrodynamic models for viscous instabilities in planetary rings. However, if diffusion and viscosity indeed arise from small-scale streaming instability, other dependencies are conceivable, namely relative dust-gas streaming velocity which powers the streaming instability in the first place.


We also neglect particle self-gravity as filaments already form before self-gravity is turned on in simulations \citep[e.g.,][]{Schreiber2018, Gerbig2020, Gerbig2023}. Nonetheless, self-gravity may have importance, modifying instability criteria and growth rates. To first order, it has a destabilizing effect and would amplify density enhancements, as well as permit gravitational instabilities \citep[e.g.,][]{Youdin2002, Youdin2011, Tominaga2019, Gerbig2020, Gerbig2023, Tominaga2023}. We reserve this topic for subsequent studies.


Lastly, we also ignored the possibility of a polydisperse dust fluid with a distribution of stopping times, and instead asserted a single stopping time, which is reasonable for a top-heavy size distribution \citep[e.g.][]{Birnstiel2012}. Still, the damping effect of particle size distributions on streaming instability \citep{Krapp2019, Paardekooper2020, Zhu2021, Yang2021} underscores that this is a relevant point to keep in mind in future investigations of diffusive instabilities. 


\section{Summary}
\label{sect:summary}

In this work we present a novel, vertically averaged axisymmetric hydrodynamic model for a dense particle layer embedded in a gaseous protoplanetary disk. The dust being dominant, we model the effect of gas as a perturbation on the dust dynamics by evoking drag forces, mass diffusion, viscosity, and pressure of the dust. The pressure is assumed to depend on the dust diffusivity following a sedimentation-diffusion ansatz, and diffusivity and viscosity are allowed to depend on the dust-surface mass density. We find that our model supports a variety of linear diffusion and oscillatory instabilities.

The diffusion instabilities arise if the dust particles' diffusion and/or viscosity decrease sufficiently fast with increasing particle surface mass density, which is motivated by the results of past simulations. Specifically, for well-coupled particles with  $\tau_\mathrm{s} \ll 1$, the diffusion-dependent pressure can destabilize the particle flow if the mass diffusion slope with respect to dust surface mass density is sufficiently negative. On the other hand, for decoupled particles $\tau_\mathrm{s} \gtrsim 1$, instability is driven by the viscosity slope, similar to the viscous instability in planetary rings.

The main application of our model is a dense midplane particle layer subject to turbulence generated by the streaming instability on small-scales.  
Indeed, the diffusivities associated with streaming instability turbulence measured in past simulations are found to be sufficient for the diffusive instabilities in our model to possess appreciable growth rates on the order of the dynamical time scale, and radial length scales that are characteristic of over-dense particle filaments seen in numerical simulations of the streaming instability. Based on these findings we argue that diffusive instabilities as captured by our model may play a role in filament formation within dusty protoplanetary disks, which is a key step within the streaming instability paradigm of planetesimal formation.

In addition, our model can also give rise to growing oscillatory modes in the presence of a sufficiently flat, or positive, viscosity slope that provides the necessary repellent acceleration.
This overstability behaves similar to the axisymmetric viscous overstability in planetary rings. Whether or not it has applicability in protoplanetary disks is unclear, as it relies strongly on the particles possessing large stopping times $\tau_\mathrm{s}>1$, as well as their interaction with the gas turbulence.

More detailed analytical investigations including vertical motions and an explicit inclusion of gas within a two-fluid formalism, accompanied by additional numerical constraints on diffusivity, viscosity, as well as their slopes,  will be crucial to further pinpoint the relevance of diffusive instabilities in dusty protoplanetary disk, filament formation therein, and planetesimal formation.

\section*{Acknowledgements}
\label{sec:acknowledgements}

\new{We are appreciative to the reviewer, whose discerning feedback was invaluable in refining the paper.} KG thanks Rixin Li, Ryosuke Tominaga, Tiger Lu and Greg Laughlin for efficacious discussions. This work is supported by the National Science and Technology Council through grants 112-2112-M-001-064- and 112-2124-M-002-003- and through an Academia Sinica Career Development Award (AS-CDA-110-M06). 

\software{ NumPy \citep{Harris2020}, Matplotlib \citep{Hunter2007}, CMasher \citep{vanderVelden2020}}.

\appendix

\section{Justification for chosen hydrodynamic equations}
\label{sect:reynoldsavering}

We utilize Reynolds averaging to justify the basic equations used in our model. Hereby, we decompose the instantaneous physical variable $A$ into average $\langle A\rangle$ and short-term fluctuation $\Delta A$ with the property $\langle \Delta A \rangle = 0$. 

Consider the continuity equation for surface density, 
\begin{align}
    \frac{\partial \Sigma}{\partial t} + \frac{1}{r}\frac{\partial (r\Sigma v_r)}{\partial r} = 0,
\end{align}
and the radial and azimuthal momentum equations
\begin{align}
\label{eq:appendix_radmoment_1}
    \frac{\partial (\Sigma v_r)}{\partial t} + \frac{1}{r}\frac{\partial }{\partial r}(r\Sigma v_r^2) = \Sigma \frac{v_\phi^2}{r} + \Sigma \Omega^2 r - \Sigma \frac{v_r - u_r}{t_\mathrm{s}}, \\
    \label{eq:appendix_phimoment_1}
    \frac{\partial (\Sigma v_\phi)}{\partial t} + \frac{1}{r}\frac{\partial }{\partial r}(r \Sigma v_\phi v_r) = -\Sigma \frac{v_\phi v_r}{r} - \Sigma \frac{v_\phi - u_\phi}{t_\mathrm{s}},
\end{align}
respectively, for an axisymmetric dust disk. The left hand side of both momentum equations is the rate of change of the momentum expressed as the sum of Eulerian derivative and advection term. The right hand side of Eq.~\eqref{eq:appendix_radmoment_1} includes curvature term, external gravitational potential and drag term. The right hand side of Eq.~\eqref{eq:appendix_phimoment_1} includes curvature term and drag term only. Reynolds decomposition and subsequent averaging yields \citep[compare to e.g.][]{Cuzzi1993, Tominaga2019}
\begin{align}
    \frac{\partial \langle \Sigma \rangle}{\partial t} & + \frac{1}{r} \frac{\partial (r \langle \Sigma \rangle \langle v_r\rangle)}{\partial r} = - \frac{1}{r}\frac{\partial}{\partial r} r\langle \Delta \Sigma \Delta v_r \rangle, \\
    \begin{split}
    \langle \Sigma \rangle \frac{\partial \langle v_r \rangle}{\partial t} & + \frac{\partial }{\partial t}\langle \Delta \Sigma \Delta v_r \rangle + \langle \Sigma \rangle \langle v_r\rangle \frac{\partial \langle v_r \rangle}{\partial r} = \langle\Sigma\rangle \frac{\langle v_\phi \rangle^2}{r} + \langle \Sigma\rangle\Omega^2 r - \langle \Sigma \rangle \frac{\langle v_r\rangle - \langle u_r\rangle}{t_\mathrm{s}} - \frac{\langle \Delta\Sigma (\Delta v_r - \Delta u_r )\rangle}{t_\mathrm{s}} \\ 
    &+ \frac{1}{r}\frac{\partial (r \sigma_{rr})}{\partial r} - \frac{\sigma_{\phi\phi}}{r} + \frac{2\langle \Delta \Sigma \Delta v_\phi \rangle \langle v_\phi\rangle}{r} - 2 \langle \Delta \Sigma \Delta v_r\rangle \frac{\partial \langle v_r \rangle }{\partial r}  - \langle v_r\rangle \frac{1}{r} \frac{\partial}{\partial r}(r \langle \Delta \Sigma \Delta v_r \rangle),
\end{split}\\
\begin{split}
    \langle \Sigma \rangle \frac{\partial \langle v_\phi \rangle}{\partial t} & + \frac{\partial}{\partial t} \langle \Delta \Sigma \Delta v_\phi \rangle  + \langle \Sigma \rangle \langle v_r \rangle \frac{\partial \langle v_\phi \rangle}{\partial r} = - \langle \Sigma \rangle \frac{\langle v_r\rangle \langle v_\phi \rangle}{r} - \langle \Sigma \rangle \frac{\langle v_\phi\rangle - \langle u_\phi \rangle}{t_\mathrm{s}} - \frac{\langle \Delta \Sigma (\Delta v_\phi - \Delta u_\phi)\rangle}{t_\mathrm{s}} \\ 
    & - \langle \Delta \Sigma \Delta v_r \rangle \frac{\langle v_\phi\rangle}{r} - \frac{1}{r}\frac{\partial }{\partial r}\left(r\langle \Delta \Sigma \Delta v_\phi\rangle \langle v_r \rangle\right) - \langle \Delta \Sigma \Delta v_r \rangle \frac{\partial \langle v_\phi \rangle}{\partial r} -  \langle \Delta \Sigma \Delta v_\phi \rangle \frac{\langle v_r \rangle}{r} + \frac{1}{r}\frac{\partial}{\partial r}(r \sigma_{r\phi}) + \frac{\sigma_{r\phi}}{r},
\end{split}
\end{align}
where we introduced the components of the Reynolds stress tensor as
\begin{align}
    \sigma_{rr} &= - \langle \Sigma \rangle \langle \Delta v_r^2 \rangle - \langle \Delta \Sigma \Delta v_r^2 \rangle, \\
    \sigma_{\phi\phi} &= - \langle \Sigma \rangle \langle \Delta v_\phi^2 \rangle - \langle \Delta \Sigma \Delta v_\phi^2 \rangle, \\
    \sigma_{r\phi} &= - \langle \Sigma \rangle \langle \Delta v_r \Delta v_\phi \rangle - \langle \Delta \Sigma \Delta v_r \Delta v_\phi \rangle.
\end{align}
Following \citet{Cuzzi1993} and \citet{Tominaga2019}, we ignore the second terms on the left hand side of both momentum equations.
We also assume that the terms $t_\mathrm{s}^{-1}\langle \Delta \Sigma (\Delta v_r - \Delta u_r)\rangle$ and $t_\mathrm{s}^{-1}\langle \Delta \Sigma (\Delta v_\phi - \Delta u_\phi)\rangle$ vanish, which is the case if $\Delta v_r = \Delta u_r$ and $\Delta v_\phi = \Delta u_\phi$ as assumed in \citet{Cuzzi1993} and \citet{Tominaga2019}.

Next, we assert the following set of closure relations
\begin{align}
\label{eq:GDH_rad}
    \langle \Delta \Sigma \Delta v_r \rangle &= -D \frac{\partial \langle \Sigma \rangle}{\partial r},\\
\label{eq:GDH_phi}
    \langle \Delta \Sigma \Delta v_\phi \rangle &= - \frac{D}{r} \frac{\partial \langle \Sigma \rangle}{\partial \phi} = 0, \\
\label{eq:Trr_Tphiphi_closure}
    \langle \Delta \Sigma \Delta v_r^2 \rangle =  \langle \Delta \Sigma \Delta v_\phi^2 \rangle &= - T_{rr} = - T_{\phi\phi} , \\
\label{eq:Trphi_closure}
    \langle \Sigma \rangle \langle \Delta v_r \Delta v_\phi \rangle + \langle \Delta \Sigma \Delta v_r \Delta v_\phi \rangle & =  - T_{r\phi} = - T_{\phi r},\\
    \label{eq:soundspeed_closure}
    \langle \Delta v_r^2\rangle = \langle \Delta v_\phi^2\rangle & = c_\mathrm{d}^2. 
\end{align}
Eqs.~\eqref{eq:GDH_rad} and \eqref{eq:GDH_phi} are the gradient diffusion hypothesis \citep[see][]{Cuzzi1993, GoodmanPindor2000, Schrapler2004, Shariff2011, Huang2022, Binkert2023}.  Eq.~\eqref{eq:soundspeed_closure} defines the effective particle velocity dispersion \citep{Cuzzi1993, Tominaga2019}. Finally, Eqs.~\eqref{eq:Trr_Tphiphi_closure} and \eqref{eq:Trphi_closure} employ the Boussinesq hypothesis \citep[also see][]{Binkert2023} for the dust fluid and in the process introduce viscosity into the problem via the viscous stress tensor in Eq.~\eqref{eq:viscous_stress_tensor} --- a choice that relates it to the Reynolds stress tensor components via
\begin{align}
    \sigma_{rr} &= - \langle \Sigma \rangle c_\mathrm{d}^2 + T_{rr} = \sigma_{\phi\phi} = - \langle \Sigma \rangle c_\mathrm{d}^2 + T_{\phi \phi},\\
    \sigma_{r\phi} &= T_{r\phi} = \sigma_{\phi r} = T_{\phi r}.
\end{align}
The correlations in Eqs.~\eqref{eq:Trr_Tphiphi_closure} and \eqref{eq:Trphi_closure} are dropped in \citet{Tominaga2019}. For $\nu = 0$, our closure relations are thus identical to theirs. \citet{Huang2022} additionally drop the pressure term in Eq.~\eqref{eq:soundspeed_closure}.

The closure relations in Eqs.~\eqref{eq:GDH_rad} -~\eqref{eq:soundspeed_closure} establish a Newtonian stress-strain relation for the particle fluid, in the process removing the need for an evolution equation for the stress tensor. This becomes questionable for $\tau_\mathrm{s} > 1$, where a kinetic approach is preferred over this fluid dynamical treatment \citep[see][and Sect.~\ref{sect:applicability}]{Jaquet2011}.

Using the gradient diffusion hypothesis, the continuity equation can directly be rewritten as an advection-diffusion equation, i.e.
\begin{align}
    \frac{\partial \langle \Sigma \rangle}{\partial t} + \frac{1}{r} \frac{\partial (r \langle \Sigma \rangle \langle v_r\rangle)}{\partial r} = \frac{1}{r}\frac{\partial}{\partial r} \left(rD\frac{\partial \langle \Sigma\rangle}{\partial r}\right).
\end{align}
In the momentum equations, all terms containing $\langle \Delta \Sigma \Delta v_\phi \rangle$ drop due to the axisymmetry. We are left with
\begin{align}
    \langle \Sigma \rangle \frac{\partial \langle v_r \rangle}{\partial t} &+ \langle \Sigma \rangle \langle v_r\rangle \frac{\partial \langle v_r \rangle}{\partial r} = \langle\Sigma\rangle \frac{\langle v_\phi \rangle^2}{r} + \langle \Sigma\rangle\Omega^2 r - \langle \Sigma \rangle \frac{\langle v_r\rangle - \langle u_r\rangle}{t_\mathrm{s}} - \frac{\partial (c_\mathrm{d}^2 \langle \Sigma\rangle)}{\partial r} + \langle \Sigma \rangle \tilde{F}_r, \\
    \langle \Sigma \rangle \frac{\partial \langle v_\phi \rangle}{\partial t} &+ \langle \Sigma \rangle \langle v_r \rangle \frac{\partial \langle v_\phi \rangle}{\partial r} = - \langle \Sigma \rangle \frac{\langle v_r\rangle \langle v_\phi \rangle}{r} - \langle \Sigma \rangle \frac{\langle v_\phi\rangle - \langle u_\phi \rangle}{t_\mathrm{s}}  + \langle \Sigma \rangle \tilde{F}_\phi,
\end{align}
where we defined
\begin{align}
    \langle \Sigma \rangle \tilde{F}_r &= \frac{1}{r}\frac{\partial}{\partial r}(r T_{rr}) - \frac{T_{\phi \phi}}{r} + \frac{1}{r}\frac{\partial}{\partial r}\left(r D \langle v_r\rangle \frac{\partial \langle \Sigma\rangle}{\partial r}\right) + D \frac{\partial \langle \Sigma \rangle}{\partial r} \frac{\partial \langle v_r \rangle }{\partial r}. \\
    \langle \Sigma \rangle \tilde{F}_\phi &=  \frac{1}{r}\frac{\partial}{\partial r}(r T_{r\phi}) + \frac{T_{r\phi}}{r} + \frac{\langle v_\phi \rangle}{r} D \frac{\partial \langle \Sigma \rangle}{\partial r} + D \frac{\partial \langle \Sigma \rangle}{\partial r}   \frac{\partial \langle v_\phi \rangle }{\partial r}  .
\end{align}
The first two terms in both equations are just the radial and azimuthal component of the divergence (in cylindrical coordinates) of the viscous stress tensor, respectively, i.e.
\begin{align}
   F_r &= \frac{1}{\langle \Sigma \rangle}(\nabla \cdot T_{ij})_r =  \frac{1}{\langle \Sigma \rangle} \left( \frac{1}{r}\frac{\partial}{\partial r}(r T_{rr}) - \frac{T_{\phi \phi}}{r} \right), \\
   F_\phi &= \frac{1}{\langle \Sigma \rangle} (\nabla \cdot T_{ij})_\phi = \frac{1}{\langle \Sigma \rangle} \left(  \frac{1}{r}\frac{\partial}{\partial r}(r \sigma_{r\phi}) + \frac{\sigma_{r\phi}}{r} \right)
\end{align}
The remaining terms in  $ \langle \Sigma \rangle \tilde{F}_r$ and $ \langle \Sigma \rangle \tilde{F}_\phi$ were already derived by \citet{Tominaga2019} and are associated with bulk transport of momentum by the diffusion flux as well as diffusive transport of bulk momentum. We thus can rewrite the momentum equations into the form
\begin{align}
\begin{split}
    \langle \Sigma \rangle \frac{\partial \langle v_r \rangle}{\partial t} + \left(\langle \Sigma \rangle \langle v_r\rangle - D \frac{\partial \langle \Sigma \rangle}{\partial r}\right)\frac{\partial \langle v_r \rangle}{\partial r} = & \langle\Sigma\rangle \frac{\langle v_\phi \rangle^2}{r} + \langle \Sigma\rangle\Omega^2 r - \langle \Sigma \rangle \frac{\langle v_r\rangle - \langle u_r\rangle}{t_\mathrm{s}} - \frac{\partial (c_\mathrm{d}^2 \langle \Sigma\rangle)}{\partial r} \\
    & + \frac{1}{r}\frac{\partial}{\partial r}\left(r D \langle v_r\rangle \frac{\partial \langle \Sigma\rangle}{\partial r}\right) + \langle \Sigma \rangle {F}_r,
\end{split}\\
    \langle \Sigma \rangle \frac{\partial \langle v_\phi \rangle}{\partial t} + \left(\langle \Sigma \rangle \langle v_r\rangle - D \frac{\partial \langle \Sigma \rangle}{\partial r}\right) \frac{\partial \langle v_\phi \rangle}{\partial r} = & - \frac{\langle v_\phi \rangle}{r}\left(\langle \Sigma \rangle \langle v_r\rangle - D \frac{\partial \langle \Sigma \rangle}{\partial r}\right) - \langle \Sigma \rangle \frac{\langle v_\phi\rangle - \langle u_\phi \rangle}{t_\mathrm{s}}  + \langle \Sigma \rangle {F}_\phi.
\end{align}
Dividing by $\langle \Sigma \rangle$ leads to Eqs.~\eqref{eq:momentum_first_r} and \eqref{eq:momentum_first_phi}. For clarity we omit the brackets $\langle \rangle$ in the paper main text. Note, that unlike \citet{Klahr2021}, we specifically keep the drag term in the azimuthal momentum equation.

\section{Dust Pressure Model}
\label{sect:dustpressuremodel}

Appendix~\ref{sect:reynoldsavering} uses Reynolds averaging to derive a pressure-like force term of the form $\partial P/\partial r$ in the radial momentum equation, where $P = c_\mathrm{d}^2 \Sigma $ can be understood as the effective dust pressure, with velocity dispersion $c_\mathrm{d}$, which we allow to vary with density. 

Specifically, following \citet{Klahr2021}, we employ a sedimentation-diffusion ansatz to model the dependence of the dust velocity dispersion on the diffusivity. The heuristic argument is to compare the settling time under linear gravity at terminal velocity $t_\mathrm{set} = 1/(t_\mathrm{s}\Omega^{2})$ with the diffusion time $t_\mathrm{diff} = 1/(Dk^2)$ across a length scale $\sim 1/k$, where $D$ is the mass diffusion coefficient of dust. Similar to Brownian motion \citep{Einstein1905}, sedimentation-diffusion equilibrium\footnote{Note, that this same argument can be made in the radial direction by replacing stellar vertical gravity with the restoring force from radial epicyclic oscillations as $-\Omega^2 x$.} leads to ${D}/{t_\mathrm{s}} = {\Omega^2}/{k^2} \equiv c_\mathrm{d}^2$. The regularized expression, that does not diverge in the limit of $\tau_\mathrm{s} = 0$, requires consideration of the gas also, leading to a dust layer thickness of $H_\mathrm{d} = \sqrt{{\delta}/({\delta + \tau_\mathrm{s}})}H$ \citep[see e.g.,][]{Lin2021}.
We define the effective particle velocity dispersion via
\begin{align}
    c_\mathrm{d} \equiv \Omega H_\mathrm{d} = \sqrt{\frac{\delta}{\delta + \tau_\mathrm{s}}} c_\mathrm{s} = \sqrt{\frac{D}{D + t_\mathrm{s}c_\mathrm{s}^2}} c_\mathrm{s}.
\end{align}
As also noted by \citet{Klahr2021}, the sedimentation-diffusion velocity dispersion $c_\mathrm{d}$ and thus our closure relation, is in general different to the root mean square (RMS) velocity of the particles. The Hinze-Tchen formalism for turbulent transport neglecting orbital oscillations or other external forces \citep{tchen1947mean, hinze1959turbulence}, gives the RMS velocity as \citep[also see e.g.,][]{fan1998principles, Youdin2007, Binkert2023}
\begin{align}
    v_\mathrm{rms}^2 & = \frac{t_\mathrm{c}}{t_\mathrm{c} + t_\mathrm{s}} u_\mathrm{rms}^2
\end{align}
with gas velocity dispersion $u_\mathrm{rms}$, and correlation time of the turbulence $t_\mathrm{c}$, which connect to the gas diffusivity $D_{\mathrm{g}}$ via  $D_\mathrm{g} = t_\mathrm{c}u_\mathrm{rms}^2$. Only if $D_\mathrm{g} \sim D$ (see Eq.~\eqref{eq:diffusion_dependence}) and for $t_\mathrm{s} \gg t_\mathrm{c}$, the RMS velocity equals the velocity dispersion following from the sedimentation diffusion ansatz. While the former is fairly well grounded in numerical simulations \citep[e.g.,][]{Schreiber2018}, the latter is somewhat more ambigous. For Kolmogorov turbulence, the correlation time equals the turnover time of the largest eddies, which in protoplanetary disks equals $\Omega^{-1}$ \citep{Youdin2007}. On the other hand, \citet{Schreiber2018} find for simulations values with active streaming instability values of $t_\mathrm{c} \sim 0.1 \Omega^{-1}$. Indeed, ignoring orbital oscillations is only a good model if $t_\mathrm{s}, t_\mathrm{c} \ll 1$. For larger particles, epicylic oscillations are important, particles can decouple from the turbulence, and the velocity dispersion and thus diffusion needs to be modified \citep[see][and also Eq.~\eqref{eq:diffusion_dependence}]{Youdin2007, Youdin2011}. We neglect this effect in this work. While our pressure term does vanishes for large stopping times like the prescriptions used by e.g., \citet[][]{Youdin2011, Umurhan2020}, the diffusivity itself does not, and is instead treated as an independent parameter. This is important to be kept in mind when evaluating our model in particular in the large stopping time limit. 

\citet{Klahr2021} call $c_\mathrm{d}$ the pseudo sound speed, which is appropriate as long as $c_\mathrm{d}$ is constant. In our work, we allow the diffusivity to depend on density, and thus $c_\mathrm{d}^2 \propto D \propto \Sigma^{\beta_\mathrm{diff}}$. As a result, the dust pressure takes the form of a polytropic equation of state, i.e. $P_\mathrm{d} \propto \Sigma^{1+\beta_\mathrm{diff}}$. One can can now formally define a dust sound speed $a_\mathrm{d}$ via 
\begin{align}
    a_\mathrm{d}^2 = \frac{\partial P}{\partial \Sigma} \propto (1+\beta_\mathrm{diff})\Sigma^{\beta_\mathrm{diff}}.
\end{align}
If $\beta_D < -1$, this effective squared sound speed is negative and as a result also the associated pressure perturbations. Indeed, such a negative pressure perturbation is a necessary (but not sufficient) requirement for the diffusion-dependent pressure driven diffusive instability that may operate for $\tau_\mathrm{s} \lesssim 1$, as discussed in this paper.

\bibliography{references}{}
\bibliographystyle{aasjournal}

\end{CJK*}
\end{document}